\documentstyle[epsfig,referee]{mn}

\newif\ifAMStwofonts

\newcommand{\bvec}[1] {\mbox{\boldmath$ #1$}}
\newcommand{\be} {\begin{equation}}
\newcommand{\ee} {\end{equation}}

\ifoldfss
  \ifCUPmtlplainloaded \else
    \NewTextAlphabet{textbfit} {cmbxti10} {}
    \NewTextAlphabet{textbfss} {cmssbx10} {}
    \NewMathAlphabet{mathbfit} {cmbxti10} {} 
    \NewMathAlphabet{mathbfss} {cmssbx10} {} 
  \fi
  \ifAMStwofonts
    \ifCUPmtlplainloaded \else
      \NewSymbolFont{upmath} {eurm10}
      \NewSymbolFont{AMSa} {msam10}
      \NewMathSymbol{\upi}     {0}{upmath}{19}
      \NewMathSymbol{\umu}     {0}{upmath}{16}
      \NewMathSymbol{\upartial}{0}{upmath}{40}
      \NewMathSymbol{\leqslant}{3}{AMSa}{36}
      \NewMathSymbol{\geqslant}{3}{AMSa}{3E}

       \let\le=\leqslant
       
    \fi
  \fi
\fi 

\ifnfssone
  \newmathalphabet{\mathit}
  \addtoversion{normal}{\mathit}{cmr}{m}{it}
  \addtoversion{bold}{\mathit}{cmr}{bx}{it}
  \newmathalphabet{\mathbfit} 
  \addtoversion{normal}{\mathbfit}{cmr}{bx}{it}
  \addtoversion{bold}{\mathbfit}{cmr}{bx}{it}
  \newmathalphabet{\mathbfss} 
  \addtoversion{normal}{\mathbfss}{cmss}{bx}{n}
  \addtoversion{bold}{\mathbfss}{cmss}{bx}{n}
  \ifAMStwofonts
    \ifCUPmtlplainloaded \else
      \UseAMStwoboldmath
      \makeatletter
      \new@mathgroup\upmath@group
      \define@mathgroup\mv@normal\upmath@group{eur}{m}{n}
      \define@mathgroup\mv@bold\upmath@group{eur}{b}{n}
      \edef\UPM{\hexnumber\upmath@group}
      \new@mathgroup\amsa@group
      \define@mathgroup\mv@normal\amsa@group{msa}{m}{n}
      \define@mathgroup\mv@bold\amsa@group{msa}{m}{n}
      \edef\AMSa{\hexnumber\amsa@group}
      \makeatother
      \mathchardef\upi="0\UPM19
      \mathchardef\umu="0\UPM16
      \mathchardef\upartial="0\UPM40
      \mathchardef\leqslant="3\AMSa36
      \mathchardef\geqslant="3\AMSa3E

       \let\le=\leqslant

    \fi
  \fi
\fi 

\ifnfsstwo
  \DeclareMathAlphabet{\mathbfit}{OT1}{cmr}{bx}{it}
  \SetMathAlphabet\mathbfit{bold}{OT1}{cmr}{bx}{it}
  \DeclareMathAlphabet{\mathbfss}{OT1}{cmss}{bx}{n}
  \SetMathAlphabet\mathbfss{bold}{OT1}{cmss}{bx}{n}
  \ifAMStwofonts
    \ifCUPmtlplainloaded \else
      \DeclareSymbolFont{UPM}{U}{eur}{m}{n}
      \SetSymbolFont{UPM}{bold}{U}{eur}{b}{n}
      \DeclareSymbolFont{AMSa}{U}{msa}{m}{n}
      \DeclareMathSymbol{\upi}{0}{UPM}{"19}
      \DeclareMathSymbol{\umu}{0}{UPM}{"16}
      \DeclareMathSymbol{\upartial}{0}{UPM}{"40}
      \DeclareMathSymbol{\leqslant}{3}{AMSa}{"36}
      \DeclareMathSymbol{\geqslant}{3}{AMSa}{"3E}

       \let\le=\leqslant

    \fi
  \fi
\fi 

\ifCUPmtlplainloaded \else
  \ifAMStwofonts \else 
    \def\upi{\pi}
    \def\umu{\mu}
    \def\upartial{\partial}
  \fi
\fi

\title{On equilibrium tides in fully convective planets and stars}
\author[P. B. Ivanov, J. C. B. Papaloizou]
       {P. B. Ivanov 
\\ Astronomy Unit, School of Mathematical Sciences, Queen Mary,
University of London, UK \\
Astro Space Center of PN Lebedev Physical 
Institute, Moscow, Russia
\newauthor J. C. B. Papaloizou
\\ Astronomy Unit, School of Mathematical Sciences, Queen Mary,
University of London, UK}

\pagerange{\pageref{firstpage}--\pageref{lastpage}}
\pubyear{1994}

\begin{document}

\maketitle

\label{firstpage}

\begin{abstract}
We  consider the tidal interaction of a binary consisting of a fully
convective primary star and a relatively 
compact mass. Using a normal mode decomposition 
we calculate the evolution of the primary angular velocity and orbit for arbitrary
eccentricity. The dissipation acting on the tidal
perturbation is assumed to  result from the action of  convective turbulence
the effects of which are assumed to act through an effective  viscosity.
A novel feature of the work presented here is that, in order to take account
of the fact that there is a relaxation time,
$t_{c},$   being the turn-over time of convective eddies, associated with the process,
this is allowed to act non locally in time
producing a  dependence of the dissipation on tidal forcing frequency.
Results are expressed in terms 
of the Fourier coefficients of the tidal potential assumed periodic in time. 
We find useful analytical approximations for  these valid for sufficiently large values of 
eccentricity $e > 0.2$.

We show that in the  framework of the equilibrium tide approximation,
when the dissipative response is frequency independent, our
results are equivalent to those 
obtained  under the  often used assumption of a constant time lag between
tidal response and forcing.

We  go on to 
consider the case  when the frequency dependence of the dissipative response is
 $ \propto 1/(1+(\omega_{m,k}t_{c})^{p})$, where
$\omega_{m,k}$ is the  apparent frequency $(( {\rm pattern speed} )\times m )$
 associated
with a particular harmonic  of
the tidal forcing as viewed in the frame corotating with the primary.
We concentrate on the case $\omega_{m,k}t_{c}  >> 1$
which is thought to be 
 appropriate to many astrophysical applications. 
We study numerically and analytically the orbital evolution
of the dynamical system corresponding to different values of the parameter $p$.
We present results from which the time
to circularize  from large eccentricity can be found.

We find that 
when $p < 1$  the evolution
is similar to that  found under the constant time lag assumption.
However, the orbital evolution of the system with
$p > 1$ differs drastically. In that case the system evolves through a sequence
of  spin-orbit corotation resonances with $\Omega_{r}/\Omega_{orb}=n/2$, where $\Omega_{r}$ and
$\Omega_{orb}$ are the rotation and orbital frequencies and $n$ is an integer.  When
$p=2$ we find an analytic expression for the evolution 
of semi-major axis with time for arbitrary
eccentricity assuming that the moment of inertia of the primary is small.

We confirm the finding of Ivanov \& Papaloizou (2004)
on the basis of an impulsive
treatment of orbits of high eccentricity that equilibrium tides 
associated with the  fundamental mode 
of pulsation and dissipative processes estimated
using the usual mixing length theory of convection seem to be  too weak
to account for the orbital circularization from large eccentricities
of extra-solar planets.

Generalisations and limitations of our formalism are discussed.

\end{abstract}


\section{Introduction}

Tidal interaction between orbiting companions produces dissipation and
orbital evolution towards a circular orbit which is  synchronised
with the internal
rotations.  Such a state is attained when the system is close enough
for the tidal interaction is strong enough (for a discussion
in the case of close binary stars
 see Giuricin,  Mardirossian \& Mezzetti 1984a,b ).
The recently discovered extra-solar planets with periods
less than about $5$ days  are in near circular orbits
which is also possibly  a result of tidal interaction ( Marcy et al 2001).

Zahn (1977) isolated two different types of tidal interaction.
The first, applicable to fully convective stars or planets, 
described as the equilibrium tidal interaction, applies when all
relevant normal modes of oscillation of the tidally disturbed object have much higher
oscillation frequencies than the orbital frequency. In this case dissipation is presumed to
occur through a turbulent viscosity which would act much like an anomalous Navier Stokes viscosity.
The second interaction described as being through dynamic tides occurs when normal
modes have oscillation frequencies comparable to the orbital frequency
and their excitation cannot be neglected.

The theory of the equilibrium tide with anomalous Navier Stokes viscosity
can be related to the classical theory of tides based on the idea that the
effect of dissipation is to cause the tidal
response   to lag behind the forcing with a constant time lag
( Darwin 1879; Alexander 1973; Hut 1981).

The tidal theory based on a constant time lag  is often applied to problems where tidal interactions
are important either in the case of stellar binaries
(eg. Lai 1999; Hurley, Tout  \& Pols  2002 and references therein)
or of giant planets  orbiting close to their
central stars (eg.  Mardling \&  Lin 2002).
It may be argued that this approach may be applicable when the
dissipative process is one that acts as a standard Navier Stokes
viscosity ( see eg. Eggleton, Kiseleva \& Hut 1998; Ivanov \& Papaloizou
2004, hereafter IP,  and below). 

However, it has been noted  on theoretical grounds
that, if the assumed
viscosity results from turbulence, there  is a  natural relaxation
time associated with it namely the eddy turn over time $t_c.$
This may be longer than the orbital period
in which case the efficiency of the dissipation process should 
be reduced and that there should be  a dependence of it on the tidal
forcing frequency (see eg.  Goldreich \& Nicholson, 1977;
Goldreich \& Keeley 1977;  Zahn 1989). Tidal
evolution may differ from that found under the constant
time lag assumption as a result. The importance of this effect
for understanding the observed properties of a particular binary
system has been  stressed by Goldman \& Mazeh (1994).

In this paper we investigate this problem
in the simplest possible context.  We study tidal evolution 
under the assumption that only the fundamental quadrupole modes are important
for determining the tidal response and that their
eigen-frequency is much larger than the orbital frequency
so that an equilibrium tide approximation can be made.
However, we allow the dissipation to have a characteristic time scale
associated with it and through this produce a frequency
dependent response in the manner that has been suggested
( eg. Goldreich \& Keeley 1977)
might  be appropriate for convective turbulence.
We find then that only when the dissipative process
acts instantaneously are results equivalent to the
constant time lag assumption obtained.
In other situations, we find that there is the possibility
of maintained spin orbit resonances and accordingly that
the state of  pseudo-synchronisation during the circularization
process differs.

The plan of the paper is as follows.
In section  \ref{BEQ} we formulate the basic equations for
calculating the response to a tidal forcing potential
in terms of a normal mode expansion. We also formulate
the treatment of dissipation using  viscosity
that incorporates a 
relaxation time,
$t_{c},$ identified with 
the turn-over time of convective eddies. This allows the
dissipative process to act non locally in time
and become weaker when the tidal forcing frequency
as seen in a frame corotating with the tidally perturbed object
becomes significantly larger than $t_c^{-1}.$
We derive expressions for the rates of
change  of energy and angular momentum applicable to an
orbit with arbitrary eccentricity. 
We introduce the equilibrium tide approximation applicable
when the only important pulsation mode for the tidal
response is the fundamental quadrupole mode and the orbital frequency
is very much less than the frequency of this mode.
We show that the results are equivalent to those
under the approximation of a constant time lag between
tidal forcing and response of the perturbed object only in the special case
$t_c =0$ which corresponds to dissipation produced
through the action of a standard Navier Stokes viscosity.

In section \ref{FCOF} we develop analytic approximations for 
the Fourier coefficients of the forcing potential.
We consider the orbital evolution calculating the circularization
from large eccentricities in section \ref{ORBE}.
We  consider the case when the angular momentum associated with rotation
of the tidally perturbed object is much less than that of the
orbit so that pseudo-synchronisation is achieved rapidly.
We suppose that the frequency dependence of the dissipative response is
 $ \propto 1/(1+(\omega_{m,k}t_{c})^{p})$, where
$\omega_{m,k}$ is the  apparent frequency $(( {\rm pattern speed} )\times m )$
associated
with a particular harmonic  of
the tidal forcing viewed in a  corotating 
frame, $p$ is a parameter   and 
$\omega_{m,k}t_{c}  >> 1.$ 
We discuss when is is possible to attain spin orbit resonance
and the form pseudo-synchronisation takes as $p$ is varied.
Both numerical and analytic approaches are used.
Finally in section \ref{CONC} we summarise our results
and discuss them in the context of extra-solar planets.

\section{Basic equations} \label{BEQ}

For the case of a fully convective planet or star considered in this paper,
the theory of tidal perturbation is relatively simple. When the
body is not rotating,
the  pressure $(p)$  modes  do not
contribute significantly to the interaction due to the  large values of their 
oscillation frequencies and the gravity $(g)$ modes are absent. Therefore, only the fundamental  $(f)$ modes  contribute significantly.
\footnote{Note, that there could be also some non-standard modes like eg. the mode appearing in a model 
of Jupiter with an assumed first order phase transition between molecular and metallic Hydrogen (eg. Vorontsov, 1984), etc.}.
For a rotating object there are also inertial or $r$  modes which
may give rise to 
a complicated spectrum and associated
eigen-functions. There could be significant excitation
depending on overlap of these  with the forcing potential.
For simplicity this   spectrum is neglected in this paper but in principle
could  be significant  when  
considering the theory of the  equilibrium tide  
( see eg. Papaloizou $\&$ Pringle 1978, 1981; Ogilvie \& Lin 
2003 for a discussion). 

We also assume that the rotation rate of the star
$\Omega_{r}$ is uniform with axis of rotation directed perpendicular to the
orbital plane, and  is sufficiently small in  magnitude  when compared 
to a  characteristic value of the eigen-frequency  $\omega_{0}$ associated with the
lowest order $f$ mode with $l=2,$.
It can be shown (see, eg. IP) 
that for our purposes it suffices to take into 
account only the leading and next order terms in the  expansion of 
all variables in the small parameter $\Omega_{r}/\omega_{0}$.
Also, only the contribution of the leading quadrupole term
in the
expansion of the tidal potential in spherical harmonics  
and hence modes with $l=2,$ are considered.
From now on we call the planet or star  on which 
the tides are raised  the primary and the perturbing
companion  the secondary.

We adopt spherical polar coordinates $(r, \theta, \phi )$
with origin at the centre of mass of the primary.
Under the  above assumptions,  the
Lagrangian displacement  of a fluid element of a fully 
convective primary
can be written in the form:
\be {\bvec \xi} =\sum_{m=0, \pm 2} 
b_{m}(t){\bvec \xi}^{m}_{0}, \label{eqno1}\ee   
where ${\bvec \xi}^{m}_{0}$ are the solutions of the standard eigen-mode equation for a non-rotating star
(eg. Tassoul 1978, Christensen-Dalsgaard 1998)
\be -\omega_{0}^{2}\bvec {\xi}^{m}_{0}+{\bf C}(\bvec {\xi}^{m}_{0})=0, \label{eqno2}\ee
where $\omega_{0}$ is the eigen-frequency of the fundamental mode in the non-rotating limit and ${\bf C}$
is the standard self-adjoint operator accounting for action of pressure and self-gravity forces on perturbations
(eg. Chandrasekhar 1964, Linden-Bell $\&$ Ostriker 1967). 

It is important to note that since the displacement vector
${\bvec \xi}$ is a real quantity and the eigen-functions 
${\bvec \xi}^{m}_{0}$ have a trivial dependence on  the azimuthal
mode number $m$ and $\phi$
$\propto e^{im\phi}$,
it follows from equation (\ref{eqno1}) that the mode amplitudes 
$b_{m}$  satisfy
\be b_{m}={(b_{-m})}^{*}, \label{eqno2a}\ee
where from now on  the superscript $*$ denotes the complex conjugate.

The eigen-functions  are orthogonal in the sense of the inner product    
\be <\bvec {\xi}_{0}^{m}| \bvec {\xi}_{0}^{n}>=\int d^3x \rho 
((\bvec {\xi}_{0}^{m})^*\cdot \bvec{\xi}_{0}^{n}), \label{eqno3}\ee     
where $\rho$ is the density of the primary star, and are normalised by the standard condition:
\be <\bvec {\xi}_{0}^{m} | \bvec {\xi}_{0}^{m}>=1. \label{eqno4}\ee
They can be presented in the form
\be \bvec {\xi}_{0}^{m}=\xi(r)_{R}Y_{2m}(\theta, \phi) {\bf e}_{r}+\xi(r)_{S}(r\nabla Y_{2m}(\theta, \phi)),  \label{eqno 4}\ee
where $Y_{2m}$ are spherical harmonics and the components $\xi(r)_{R}$ and $\xi(r)_{S}$ are related to each other as
\be {d\over dr}\xi_{S}={\xi_{R}-\xi_{S}\over r}. \label{eqno 5}\ee
This condition follows from the fact that  hydrodynamical motion
induced in a non-rotating isentropic star  by a forcing potential must be
circulation free.

The equation governing the Lagrangian displacement produced under
tidal forcing is (see eg. IP)
\be {\ddot{\bvec {\xi}}}+ 2{\bf u}_0 \cdot\nabla {\dot{\bvec {\xi}}}
+ {\bf C}(\bvec {\xi})=
-\nabla \psi'+ {\bf f}^{\nu}, \label{eqno2b}\ee
where ${\bf u}_0$ is the 
rotational velocity of the primary and the viscous force per unit mass is 
${\bf f}^{\nu}.$

The evolution equation for the mode amplitudes $b_{m}$ obtained by projecting
onto the eigenvectors ${\bvec {\xi}}^m_0$
 is found to be (see IP)
\be \ddot b_{m}+\omega^{2}_{0}b_{m}+2im\beta\Omega_{r}\dot b_{m}=f^{T}_{m}+f^{\nu}_{m}. \label{eqno 6}\ee
Here the dimensionless coefficient $\beta$ determines correction to 
$\omega_{0}$ due to rotation, it has a form (eg. Christensen-Dalsgaard 1998 and references
therein): 
\be \beta=1-\int^{R_{pl}}_{0}r^{2}dr\rho(2\xi_{R}\xi_{S}+\xi_{S}^2). \label{eqno 8}\ee

The forcing amplitude $f_{m}^{T}$ is related to the forcing tidal
potential $\psi'$ (assumed to be $\propto \exp(im\phi) )$  through
\be f_{m}^{T} = -\int d^3x \rho (\bvec {\xi}_{0}^{m})^*\cdot \nabla \psi' \label{ft} \ee

For our problem, the quantity $f_{m}^{T}$   can be written in the form
\be  f_{m}^{T}={GMQ\over D(t)^{3}}W_{m}e^{-im\Phi(t)}, \label{eqno18}\ee
where the overlap integral $Q$ is expressed as (Press $\&$ Teukolsky 1977)
\be Q=2\int^{R_{pl}}_{0}dr\rho r^{3}(\xi_{R}+3\xi_{S}). \label{eqno9} \ee 
Here the mass of the secondary star is $M$, ${\bf D}(t)$ is  the position vector of
the secondary assumed to  orbit in the plane $\theta = \pi/2$ with 
$D(t)=|{\bf D}(t)|$ and $\Phi(t)$  being  the associated azimuthal angle 
at some arbitrary moment of time $t$. For $|m|=2,$
$W_m= \sqrt{3\pi/10}$ and $W_0 =-\sqrt{\pi/5}$.
We find it useful below to represent  $f_{m}^{T}$ as multiplication of a constant dimensional factor
and a time-dependent dimensionless function:
\be  f_{m}^{T}=W_{m}C\phi_{m}(t), \quad C={GMQ\over a^{3}}, 
\quad \phi_{m}(t)={e^{-im\Phi(t)}\over {\tilde D(t)}^{3}}, \label{eqno9a} \ee
where $a$ is the semi-major axis and ${\tilde D(t)}=D(t)/a$. 

\subsection* {The  viscous force  amplitudes}
The amplitude  associated with the viscous force
is given by $f^{\nu}_{m}$ in equation (\ref{eqno 6}). 
For the standard
Navier Stokes equations
 $f^{\nu}_{m}$ can be directly calculated (see IP and  below).
 However, in this paper
we  consider a more general type of viscous interaction 
presumed to be induced by  convective 
turbulence  and for which the expression for $f^{\nu}_{m}$
must be modified. In particular, as has been pointed out 
by a number of authors (eg. Goldreich $\&$ Nicholson 1977;
Goldreich $\&$ Keeley  1977),
when the characteristic tidal forcing time is much 
smaller than the characteristic turn-over time of convective eddies,
 $t_{c}$, the
transport of energy and momentum by turbulent motions is ineffective and  
the effective viscosity coefficient should be relatively suppressed.
In order
to describe 
the mode energy   dissipation arising from the turbulence,
 a time interval of extent
$t_{c}$   centred on the local time $t$
should be considered. We would like to take  this into
account  in a simple manner and accordingly
 introduce a non-local form of $f^{\nu}_{m}$  written in the form
\be f_{m}^{\nu}=-\int^{\infty}_{-\infty} dt' \gamma_{m}(t-t')(\dot b_{m}(t')
+im\Omega_{r}b_{m}(t')), \label{eqno 10} \ee
where the  fact that $\dot b_{m}+im\Omega_{r}b_{m}$ is used means that
the dissipation operates
on the amplitude of the
 velocity field associated with a particular mode  as viewed in the 
frame co-rotating with the primary.
We specify a particular form of viscosity kernel
$\gamma_{m}(t)$ below (see equation \ref{eqno 55})), and 
discuss here only its most general properties.

Since the tidal forcing time and the turn-over time must be estimated in the co-rotating frame, and modes with different $m$ have different
apparent frequencies in this frame, $\gamma_{m}(t)$ should depend both on $m$
and $\Omega_{r}$. For general linear perturbations,
$\gamma_{m}(t)$ is a 
complex quantity, but the condition
 $\gamma_{m}(t)={(\gamma_{-m}(t))}^{*}$ must be valid 
for self-consistency of equation  (\ref{eqno 6}). 

In the limit $t_{c} \rightarrow 0$ the viscous interaction is local in time and  $\gamma_{m}(t)=\gamma_{0}\delta (t)$,
where $\gamma_{0}$ is the viscous mode damping rate. It  has dimensions
 of inverse time for a normalised eigen-function.  
In that case  which corresponds to
standard Navier Stokes viscosity  it may 
be explicitly expressed in terms of the components $\xi(r)_{R}$ and $\xi(r)_{S}$ and the
density $\rho$ and kinematic viscosity $\nu$ of the star (IP):
\be \gamma_{0}=4\int^{R_{pl}}_{0} r^{2}dr\rho \nu \lbrace {1\over 3} (\xi_{R}^{'}-{\xi_{R}\over r}+
3{\xi_{S}\over r})^{2}+
{1\over r^{2}}({(\xi_{R}-3\xi_{S})}^{2}+2\xi_{R}^{2}+3{(\xi_{R}-\xi_{S})}^{2})\rbrace, 
\label{eqno 8a}\ee 
where the prime stands for differentiation with respect to $r$.

\subsection*{Energy and Angular Momentum Exchange between primary
and orbit}
Recalling that for the problem on hand only modes
with $|m| = 0, 2$ are excited
it follows from equation (\ref{eqno 6}) that 
when $f_{m}^{T}$ and $f_{m}^{\nu}$ $(|m| = 0, 2)$
 are equal to zero the following 
quantities are conserved:
\be E=\dot b_{2}{\dot b_{2}}^{*}+{{\dot b_{0}}^{2}\over 2}
+\omega^{2}_{0}(b_{2}b_{2}^{*}+{b_{0}^{2}\over 2}), \label{eqno 11}\ee
where we recall $b_0$ is real
and
\be L=2(i({b_{2}}^{*}\dot b_{2}-b_{2}{(\dot b_{2})}^{*})-4\beta \Omega_{r}b_{2}{b_{2}}^{*}).\label{eqno 120}\ee
These quantities represent the sum of the
canonical energies and angular momenta associated with 
the different modes (e. g. Friedmann $\&$ Schutz, 1978).
 
When the effects of dissipation
are included, 
the evolution equations for $E$ and $L$ follow from equation (\ref{eqno 6}) as
\be  \dot E=\dot E_{T}+\dot E_{\nu}, \quad \dot L =\dot L_{T}+\dot L_{\nu}, \label{eqno 12}\ee  
where
\be \dot E_{\alpha}=\dot b_{2}{(f_{2}^{\alpha})}^{*}+{\dot b_{2}}^{*}f^{\alpha}_{2}+\dot b_{0}f_{0}^{\alpha},
\quad \dot L_{\alpha}=2i(b_{2}^{*}f_{2}^{\alpha}-b_{2}{(f_{2}^{\alpha})}^{*}),  \label{eqno 13}\ee
where the index $\alpha$ is either $T$ or $\nu$.

The sum of  the orbital energy $E_{orb}$, the energy of 
the oscillation  modes $E$ and the energy of the primary star 
$E_{st}$ is conserved during the
evolution of the system
\footnote{For simplicity we assume hereafter that the tides are raised only in the primary star. The generalisation to the
case of two tidally interacting stars is straightforward. Also, note that the energy dissipated by tidal friction
and finally radiated away from the primary is 
formally considered as a part of $E_{st}$.},
as is the sum of the
corresponding angular momenta. 
If the slow change of orbital parameters due 
to tidal evolution is neglected , the pulsations of the primary star are
strictly  periodic  and the energy of the modes $E$ and 
the  associated angular momentum $L$ time  averaged over a period
doesn't grow with time. In this case we have
\be \langle \dot E_{orb} \rangle =-\langle \dot E_{\nu} \rangle, 
\quad \langle \dot L_{orb} \rangle =-\langle \dot L_{\nu} \rangle , \label{eqno 14}\ee
where   the angular brackets
denote a time average 
such that for any quantity $Q,$
$\langle Q \rangle={1\over P_{orb}}\int^{P_{orb}}_{0}dtQ(t).$

Note that the mean rate of energy dissipation 
$\langle \dot E_{diss} \rangle $  differs from 
$\langle \dot E_{\nu} \rangle$ for a rotating primary  being given by 
\be \langle \dot E_{diss} \rangle =\langle \dot E_{\nu} \rangle-\Omega_{r}\langle \dot L_{\nu}\rangle, \label{eqno 15}\ee
and must be negative. On the other hand the sign of $\dot E_{\nu}$ can be arbitrary.

\subsection* {Solution for the mode amplitudes}
\subsubsection{Fourier expansion of the mode and forcing amplitudes}

In order to solve equation (\ref{eqno 6}) for the mode amplitudes
we assume that the orbital parameters are fixed 
 so that all quantities of interest are periodic functions of time  with period $P_{orb}$.
Note that a care should be taken with this assumption (see IP and references therein and also the 
discussion), especially for  orbits with
high eccentricity $e$
\footnote{For a highly eccentric orbits this assumption can  break down  when
the  excitation of so-called dynamic tides is significant
(eg. Press $\&$ Teukolsky 1977).  Also the  
stochastic evolution of orbital parameters (eg. Mardling 1995, IP and references therein)
may take place.}.

We use below a Fourier expansion for the mode amplitudes $b_{m}$ and the dimensionless tidal forcing amplitudes $\phi_{m}$ 
in the form:
\be b_{m}=\sum_{k=-\infty}^{\infty}b_{m,k}e^{ik\tau}, \label{eqno 16}\ee  
\be \phi_{m}=\sum_{k=-\infty}^{\infty}\phi_{m,k}e^{ik\tau}, \label{eqno 17}\ee 
where we introduce the dimensionless time variable
$\tau=\Omega t$, and $\Omega$ is the orbital frequency:
\be \Omega=\sqrt{{G(M+M_{p})\over a^{3}}}=\sqrt{{GM(1+q)\over a^{3}}}, \label{eqno 17c}\ee
where $M_{p}$ is the mass of the primary, and $q=M_{p}/M$ is the mass ratio.
 
From the condition $b_{m}={(b_{-m})}^{*}$ it follows that
$b_{2,k}={(b_{-2,-k})}^{*}$. The amplitude $b_{0}$ is real and therefore $b_{0,k}={(b_{0,-k})}^{*}$. 

The Fourier coefficients of the dimensionless tidal forcing amplitudes $\phi_{m,k}$ are given by
\be \phi_{m,k} = {1\over 2\pi}\int^{2\pi}_0 {e^{-im\Phi(t) -ik\tau}\over \tilde D(t)^{3}} d\tau. 
\label{eqno 18} \ee
We note that $\tilde D$ and $\Phi$ are related to $\tau$ through
$\tilde D(\tau)= (1-e\cos(\xi))$ and
$\tan \Phi = \sqrt{1-e^2}\sin\xi/(\cos\xi -e )$  
with $\tau=\xi - e\sin(\xi)$.

On account of the fact that $\tilde D$ and $\Phi$  are  even function
and odd functions of $\tau$ respectively, we note that
$\phi_{m,k}$ is a function only
of the eccentricity and is real. We may therefore write
\be \phi_{m,k}  ={1\over 2}(\alpha_{m,k} - \beta_{m,k}), \label{eqno 19}\ee 
where 
\be \alpha_{m,k}={1\over \pi}\int^{2\pi}_{0}d\tau {\cos(m\Phi(\tau))cos(k\tau)\over \tilde D(\tau)^{3}}, \quad
\beta_{m,k}={1\over \pi}\int^{2\pi}_{0}d\tau {\sin(m\Phi(\tau))sin(k\tau)\over \tilde D(\tau)^{3}}. \label{eqno 20} \ee
We note  obvious properties of the coefficients 
such as $\alpha_{m,-k} = \alpha_{m,k}$, $\beta_{m,-k} = -\beta_{m,k}$ 
and $\alpha_{-m,k} = \alpha_{m,k}$, $\beta_{-m,k} = -\beta_{m,k}$.
Furthermore it is important to note that for the case of interest with
$m=2$ and positive values of $k$ we approximately have  
$\alpha_{2,k}\approx \beta_{2,k}$, and for $m=0,$ $\beta_{0,k}=0$.

We find approximate expressions   for 
$\alpha_{2,k}$ and $\alpha_{0,k}$  for positive values of $k,$
these only being required,  in the  Appendix.
 We find  it convenient below to represent
the dimensionless tidal forcing amplitudes $\phi_{\pm2}$ in terms of the  real quantities
$(\phi_{+}, \phi_{-})$  defined through
\be \phi_{\pm2}=\phi_{+}\mp i\phi_{-}. \label{eqno 21} \ee 
One can easily check that $\alpha_{2,k}$ ($\beta_{2,k}$) are  the  Fourier cosine (sine)  expansion 
coefficients of $\phi_{+}$
($\phi_{-}$), and $\alpha_{0,k}$ are  the Fourier cosine  expansion  coefficients 
of $\phi_{0}$.

We introduce the Fourier transform of the kernel function
\be \gamma_{m,k} =\int^{\infty}_{-\infty} d\tau \gamma_{m}(\tau )\exp (-ik\tau )\ee
It is convenient
to work with a dimensionless form  ${\tilde \gamma_{m,k}}$ defined through 
\be \gamma_{m,k}= \bar \gamma \tilde \gamma_{m,k} , \label{eqno 22} \ee 
where $\bar \gamma$ is some characteristic value of the damping rate.  
For the
case of the standard Navier Stokes  viscosity, $t_c \rightarrow 0,$
$\bar \gamma = \gamma_{0}$ and $\tilde \gamma_{m,k}=1$. 

In order to find solutions of equation 
(\ref{eqno 6}) we first  express $f^{\nu}_{m}$ in terms of the
$\gamma_{m,k}$ and $b_{m,k}$ with
the help of equations (\ref{eqno 10}) and (\ref{eqno 16})
so obtaining 
\be f^{\nu}_{m}=-i\Omega \sum_{-\infty}^{\infty} (k+m\sigma)\gamma_{m,k}b_{m,k}e^{ik\tau},
\label{eqno 23}\ee
where 
\be \sigma={\Omega_{r}\over \Omega}. \label{eqno 23a}\ee
After substitution of equations  (\ref{eqno 16}-\ref{eqno 17}) and equation
 (\ref{eqno 23}) into equation  (\ref{eqno 6}), we obtain
\be b_{m,k}={W_{m}C\phi_{m,k}\over R}, \label{eqno 24}\ee
with 
\be R=\omega_{0}^{2}-(k^{2}\Omega + m\beta \Omega_{r} -i(k+2m\sigma)
\gamma_{m,k})\Omega. \label{eqno 25}\ee

We obtain expressions for $\langle \dot E_{\nu} \rangle$ 
and $\langle \dot L_{\nu} \rangle$ by substituting (\ref{eqno 16}) and (\ref{eqno 23})
in equations (\ref{eqno 12} - \ref{eqno 13})
and averaging the result over a period $2\pi$ in  $\tau$
so obtaining 
\be \langle \dot E_{\nu} \rangle =-2\Omega^{2} \sum_{1}^{\infty}\lbrace k(k-2\sigma) \gamma_{2,-k}b_{2,-k}b_{2,-k}^{*}+
k(k+2\sigma)\gamma_{2,k}b_{2,k}b_{2,k}^{*}+k^{2}\gamma_{0,k}b_{0,k}b_{0,k}^{*}\rbrace, 
\label{eqno 26}\ee
\be \langle \dot L_{\nu} \rangle =-4\Omega \sum_{1}^{\infty}\lbrace (k-2\sigma)\gamma_{2,-k}b_{2,-k}b_{2,-k}^{*}-
(k+2\sigma)\gamma_{2,k}b_{2,k}b_{2,k}^{*}\rbrace, \label{eqno 27}\ee
where we have used the symmetry properties of $b_{m,k}$ and $\gamma_{m,k}$
as well as the fact that the Fourier coefficients of the potential
vanish when $k=0$
and $m=2,$ to obtain a sum over positive integers 
$k$ only. It is easy to see that
\be \langle \dot E_{diss} \rangle =
-2\Omega^{2}\sum_{1}^{\infty}\lbrace (k+2\sigma)^{2}
\gamma_{2,k}b_{2,k}b_{2,k}^{*}+
(k-2\sigma)^{2}\gamma_{2,-k}b_{2,-k}b_{2,-k}^{*}+k^{2}\gamma_{0,k}b_{0,k}b_{0,k}^{*}\rbrace \label{eqno 28}\ee
is always negative when $\gamma_{m,k} > 0$. 

Equations  (\ref{eqno 26} - \ref{eqno 28}) and (\ref{eqno 24})  
together with expressions (\ref{eqno 18}) for $\phi_{m,k}$ and the standard 
Keplerian expressions for the orbital energy and angular momentum give a complete set
of equation for  our model.

\subsection*{The equilibrium tide limit}

\subsubsection{General energy and angular momentum exchange rates}
In the approximation scheme corresponding to calculation of the
equilibrium (or quasi-static) tide, the orbital and rotational frequencies and the mode damping rate are 
very much smaller in magnitude than the oscillation frequency $\omega_0$.
In this case these can be neglected in the denominator $R$ of equation (\ref {eqno 24}).
The Fourier mode amplitudes $b_{m,k}$ are then  simply proportional to $\phi_{m,k}$ such that
\be b_{m,k}={W_{m}C\phi_{m,k}\over \omega_{0}^{2}}.  \label{eqno 29} \ee

Substituting (\ref{eqno 29}) in (\ref{eqno 26}) and (\ref{eqno 27}) we obtain
\be \langle \dot E_{\nu} \rangle =-S_{1}\dot E_{\nu}^{0}, \label{eqno 30}\ee 
\be \langle \dot L_{\nu} \rangle =-S_{2}\dot L_{\nu}^{0}, \label{eqno 31}\ee 
where
\be \dot E_{\nu}^{0}={3\pi \over 10} {\Omega^{2} C^{2}\bar \gamma \over \omega_{0}^{4}}, \quad 
\dot L_{\nu}^{0}={3\pi \over 5} {\Omega C^{2}\bar \gamma \over \omega_{0}^{4}}, \label{eqno 32}\ee
and
\be S_{1}=2\sum_{1}^{\infty} \lbrace k(k-2\sigma) 
\tilde \gamma_{2,-k}\phi_{2,-k}^{2}+
k(k+2\sigma)\tilde \gamma_{2,k}
\phi_{2,k}^{2}+{2\over 3}k^{2}\tilde \gamma_{0,k}\phi_{0,k}^{2} 
\rbrace, \label{eqno 30c}\ee  
\be S_{2}=2\sum_{1}^{\infty} 
\lbrace (k-2\sigma)\tilde \gamma_{2,-k}\phi_{2,-k}^{2}-
(k+2\sigma)\gamma_{2,k}\phi_{2,k}^{2}\rbrace. \label{eqno 31c}\ee 
Here we use the definitions of $\bar \gamma$ and $\tilde \gamma_{m,k}$, the explicit form of $W_{m}$ and the fact that $\phi_{m,k}$ are
real. The quantities $\dot E^{\nu}_{0}$ and $\dot L^{\nu}_{0}$ provide  characteristic values 
of  the rates of change of the energy and angular momentum
and set  a characteristic time scale of tidal evolution.
Taking into account the definition of $C$ (equation (\ref{eqno9a})) and the normalisation condition
for the eigen-modes (\ref{eqno4}) one can easily see that 
these quantities have the correct dimensions of energy and angular momentum per
unit of time respectively.

As discussed above,
 $|\phi_{2,k}|\ll |\phi_{2,-k}|$ and in a good approximation we can neglect the terms proportional to
$\phi_{2,k}^{2}$ in equations  (\ref{eqno 30}-\ref{eqno 31c}) and rewrite them in a simpler form:
\be \langle \dot E_{\nu} \rangle\approx -2 \dot E_{\nu}^{0}\sum_{1}^{\infty} \lbrace k(k-2\sigma) \tilde \gamma_{2,-k}\phi_{2,-k}^{2}+
{2\over 3}k^{2}\tilde \gamma_{0,k}\phi_{0,k}^{2} \rbrace, \label{eqno 33a}\ee 
\be \langle \dot L_{\nu} \rangle \approx -2\dot L_{\nu}^{0}\sum_{1}^{\infty}(k-2\sigma)\tilde \gamma_{2,-k}\phi_{2,-k}^{2}. \label{eqno 34a}\ee   
We use equations  (\ref{eqno 33a})  and (\ref{eqno 34a}) below for our numerical calculations.

\subsubsection{ An alternative form}

There is a form of equations  (\ref{eqno 30}-\ref{eqno 31c}) 
which is useful for relating the work here to that 
of Hut (1981)  in  which he studied  tidal interactions of  general orbits
using   a constant  response 
time lag approximation together with the  assumption of equilibrium tides.
We write them in the form:

\be \langle \dot E_{\nu} \rangle = -\dot E_{\nu}^{0}(\Psi_{1}-2\sigma \Psi_{2}), \label{eqno 35}\ee
\be \langle \dot L_{\nu} \rangle =-\dot L_{\nu}^{0}(\Psi_{2}-2\sigma \Psi_{3}), \label{eqno 36}\ee
where
\be \Psi_{1}={1\over 2} \sum_{1}^{\infty}k^{2}\lbrace \tilde \gamma_{2,-k}(\alpha_{2,k}+\beta_{2,k})^{2}+
\tilde \gamma_{2,k}(\alpha_{2,k}-\beta_{2,k})^{2}+{2\over 3}\tilde \gamma_{0,k}\alpha_{0,k}^{2}\rbrace, \label{eqno 37}\ee
\be \Psi_{2}={1\over 2}\sum_{1}^{\infty}k \lbrace \tilde \gamma_{2,-k}(\alpha_{2,k}+\beta_{2,k})^{2}-
\tilde \gamma_{2,k}(\alpha_{2,k}-\beta_{2,k})^{2}\rbrace, \label{eqno 38}\ee
\be \Psi_{3}={1\over 2} \sum_{1}^{\infty}\lbrace \tilde \gamma_{2,-k}(\alpha_{2,k}+\beta_{2,k})^{2}+
\tilde \gamma_{2,k}(\alpha_{2,k}-\beta_{2,k})^{2}\rbrace. \label{eqno 39}\ee 
Here we  have 
expressed  $\phi_{2,\pm k}$ in terms of $\alpha_{2,k}$
and $\beta_{2,k}$ using equation (\ref{eqno 19}). Obviously, we have 
$S_{1}=\Psi_{1}-2\sigma \Psi_{2}$ 
and $S_{2}=\Psi_{2}-2\sigma \Psi_{3}$.

\subsubsection{The limit $t_c \rightarrow 0$ }

The limit $t_c \rightarrow 0$   corresponds 
to local behaviour and a Navier Stokes viscosity.
Assuming in this limit  that 
$\gamma_{m}(t)=\gamma_{0}\delta (t),$ we may adopt
 $\tilde \gamma_{m,k}=1$ in equations (\ref{eqno 37}-\ref{eqno 39}). 
In this case the infinite series (\ref{eqno 37}-\ref{eqno 39}) 
can be exactly summed by use of arguments exploiting  Parseval's theorem. 

Then we directly  obtain from  equations (\ref{eqno 37}-\ref{eqno 39}) that
\be \Psi_{1}^{0}=\sum_{1}^{\infty}k^{2}(\alpha_{2,k}^{2}+\beta_{2,k}^{2}+{1\over 3}\alpha_{0,k}^{2}), \label{eqno 40}\ee
\be \Psi_{2}^{0}=2\sum_{1}^{\infty}k \alpha_{2,k}\beta_{2,k}, \label{eqno 41}\ee
\be \Psi_{3}^{0}=\sum_{1}^{\infty}(\alpha_{2,k}^{2}+\beta_{2,k}^{2}). \label{eqno 42}\ee 
Using the definitions of $\alpha_{m,k}$ and $\beta_{m,k}$ (equation (\ref{eqno 20})) and also of $\phi_{+}$, $\phi_{-}$ and $\phi_{0}$ 
(equation (\ref{eqno 21}), one can easily show that the series (\ref{eqno 40}-\ref{eqno 42}) can be expressed in terms of integrals:
\be \Psi_{1}^{0}={1\over \pi}\int^{2\pi}_{0}d\tau ({({d\phi_{+}\over d\tau})}^{2}+{({d\phi_{-}\over d\tau})}^{2}+
{1\over 3}{({d\phi_{0}\over d\tau})}^{2}), \label{eqno 43}\ee
\be \Psi_{2}^{0}={1\over \pi}\int^{2\pi}_{0}d\tau (\phi_{+}{d \phi_{-}\over d\tau}-\phi_{-}{d \phi_{+}\over d\tau}), \label{eqno 44}\ee
\be \Psi_{3}^{0}={1\over \pi}\int^{2\pi}_{0}d\tau (\phi_{+}^{2}+\phi_{-}^{2}). \label{eqno 450}\ee  
The evaluation of the integrals (\ref{eqno 43}-\ref{eqno 44}) is straightforward with the result:
\be \Psi_{1}^{0}={8\over \epsilon^{15}}(1+{31\over 2}e^{2}+{255\over 8}e^{4}+{185\over 16}e^{6}+{25\over 64}e^{8}), \label{eqno 45}\ee
\be \Psi_{2}^{0}={4\over \epsilon^{12}}(1+{15\over 2}e^{2}+{45\over 8}e^{4}+{5\over 16}e^{6}), \label{eqno 46}\ee
\be \Psi_{3}^{0}={2\over \epsilon^{9}}(1+3e^{2}+{3\over 8}e^{4}), \label{eqno 47}\ee
where
$\epsilon=\sqrt{(1-e^{2})}$. 

When the expressions (\ref{eqno 45}-\ref{eqno 47}) are inserted into
equations (\ref{eqno 35} - \ref{eqno 36}) the
resulting exchange rates of energy and angular momentum
are identical to
the corresponding expressions obtained by Hut (1981) 
provided we make the  identification
\be {k_{H}\over T_{H}}={4\pi\over 5}{\bar \gamma M_{p}(GQ)^{2}\over \omega_{0}^{4}R_{p}^{8}}, \label{eqno 47c} \ee
where $R_{p}$ is the primary radius, $k_{H}$ is the apsidal motion constant defined as in Hut (1981), and
$T_{H}$ is Hut's typical time of tidal evolution.

Thus, in the limit 
of constant (frequency independent) 
viscosity, our normal mode approach to the problem is equivalent to the 
standard constant time lag approach (Alexander, 1973,
Hut, 1981). However, it has the advantage of enabling
us to express the 
unspecified evolution time $T_{H}$ in terms of micro-physical quantities
such as the kinematic viscosity $\nu$ and the primary density $\rho$
and quantities related to the normal mode 
(see also Eggleton, Kiseleva \& Hut 1998).
Note that equation
(\ref{eqno 47c}) has been obtained by IP 
in the limit of highly eccentric orbits
\footnote{Note a misprint in IP. Their coefficients $\beta_{1}$ and $\beta_{2}$, and accordingly the numerical factor
entering their equation (46) 
analogous to equation (\ref{eqno 47c}) must be divided on two.} 
(see also equations ( 23 - 25 ) of Kumar, Ao $\&$ Quataert 1995).

\section{The  Fourier expansion of the tidal force amplitude} \label{FCOF}

\begin{figure*}
\vspace{8cm}\includegraphics{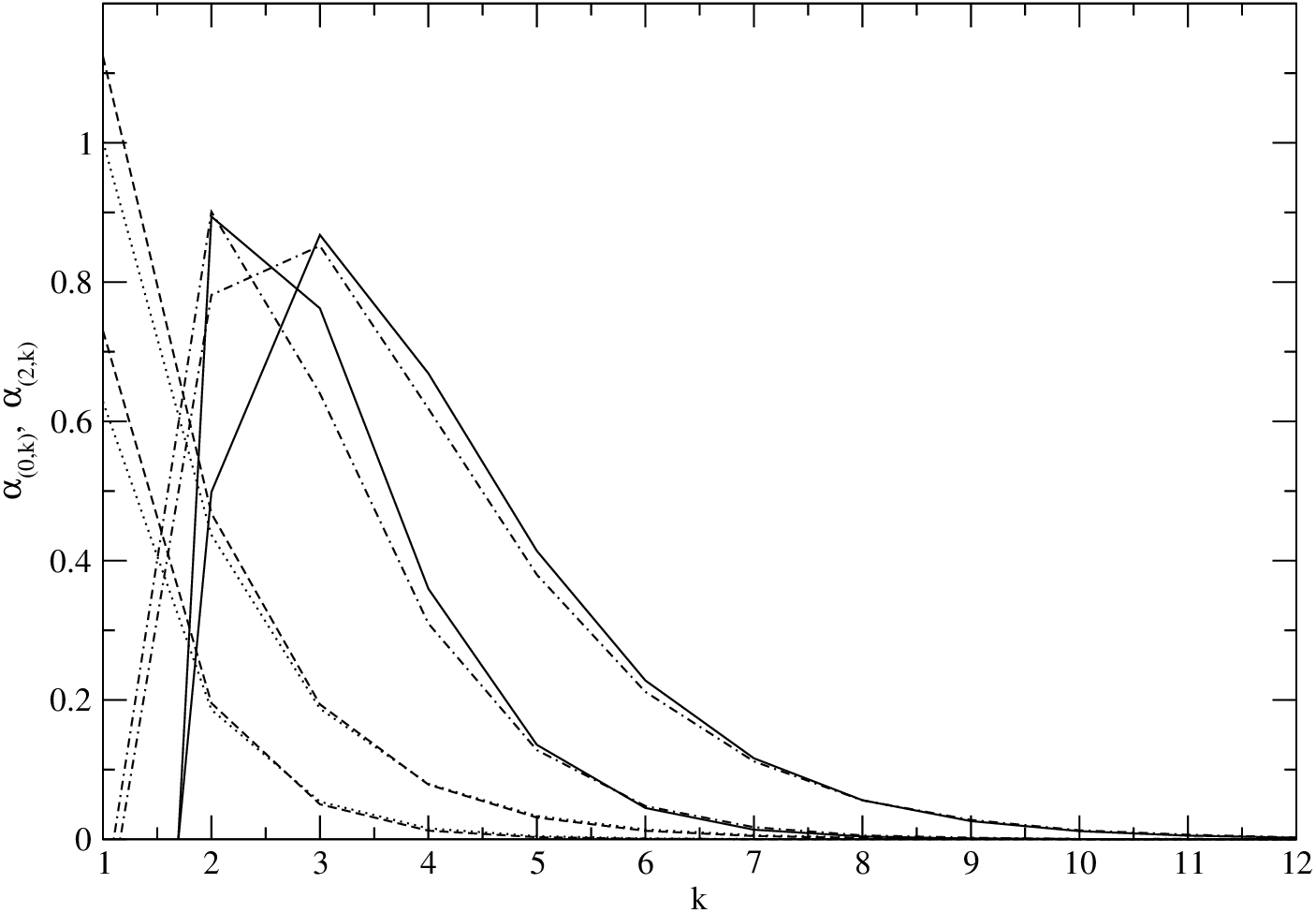}
\caption{The coefficients $\alpha_{0,k}$ (dashed lines)
 and $\alpha_{2,k}$ (solid lines) obtained with the analytic
approximations  together with the same quantities 
obtained numerically (dotted and dot-dashed lines respectively). 
 Curves of the same
type with smaller values of $\alpha_{m,k}$ correspond to $e=0.2$ and 
those with the larger value to $e=0.3$. \label{fig1}}
\end{figure*}

\begin{figure*}
\vspace{8cm}\includegraphics{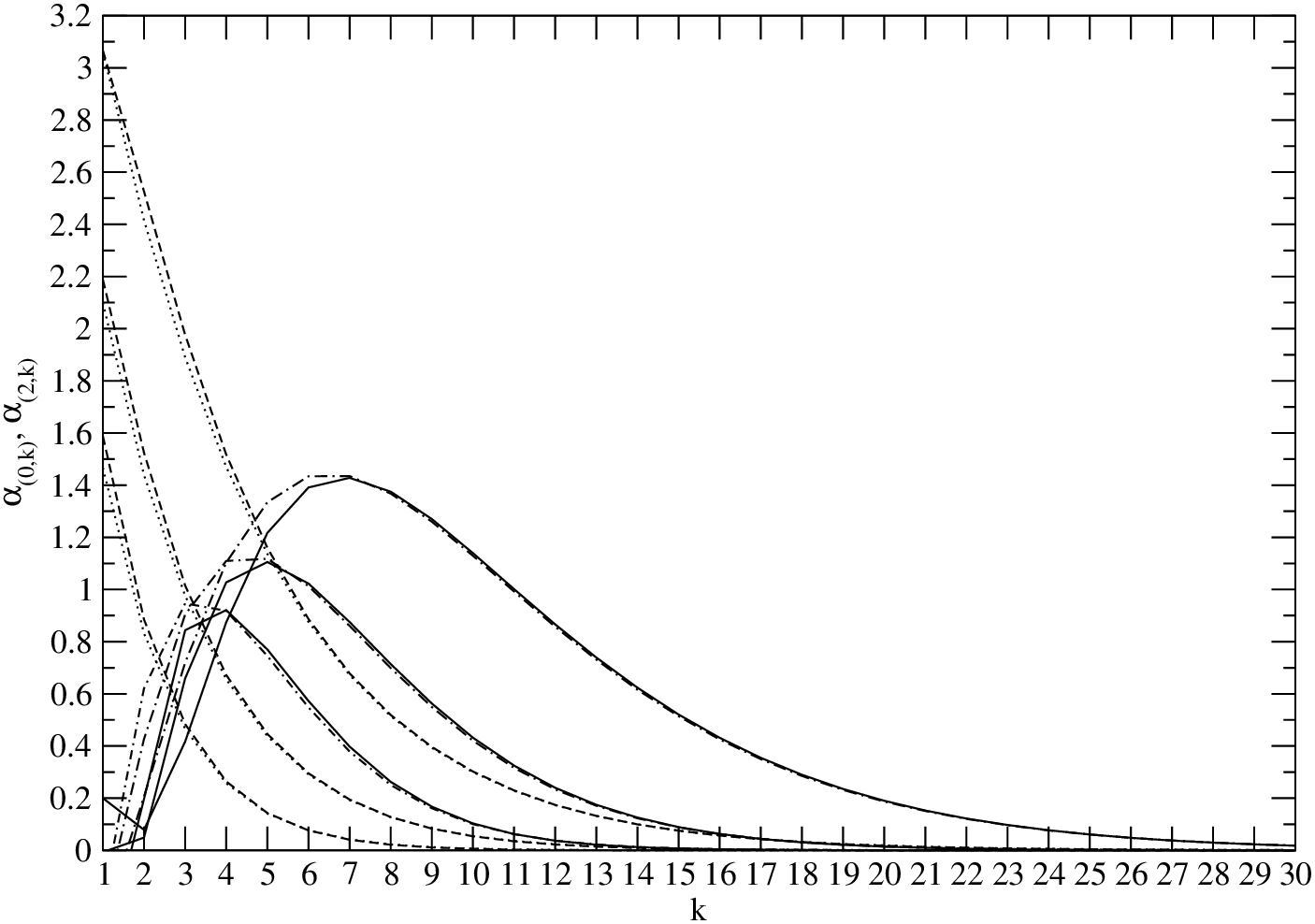}
\caption{As in \ref{fig1}
 but for $e=0.4$, $0.5$  and $0.6.$ Curves of the same type with  
increasing values of $\alpha_{m,k}$
correspond to increasing values of $e.$ \label{fig2}}
\end{figure*}

\begin{figure*}
\vspace{8cm}\includegraphics{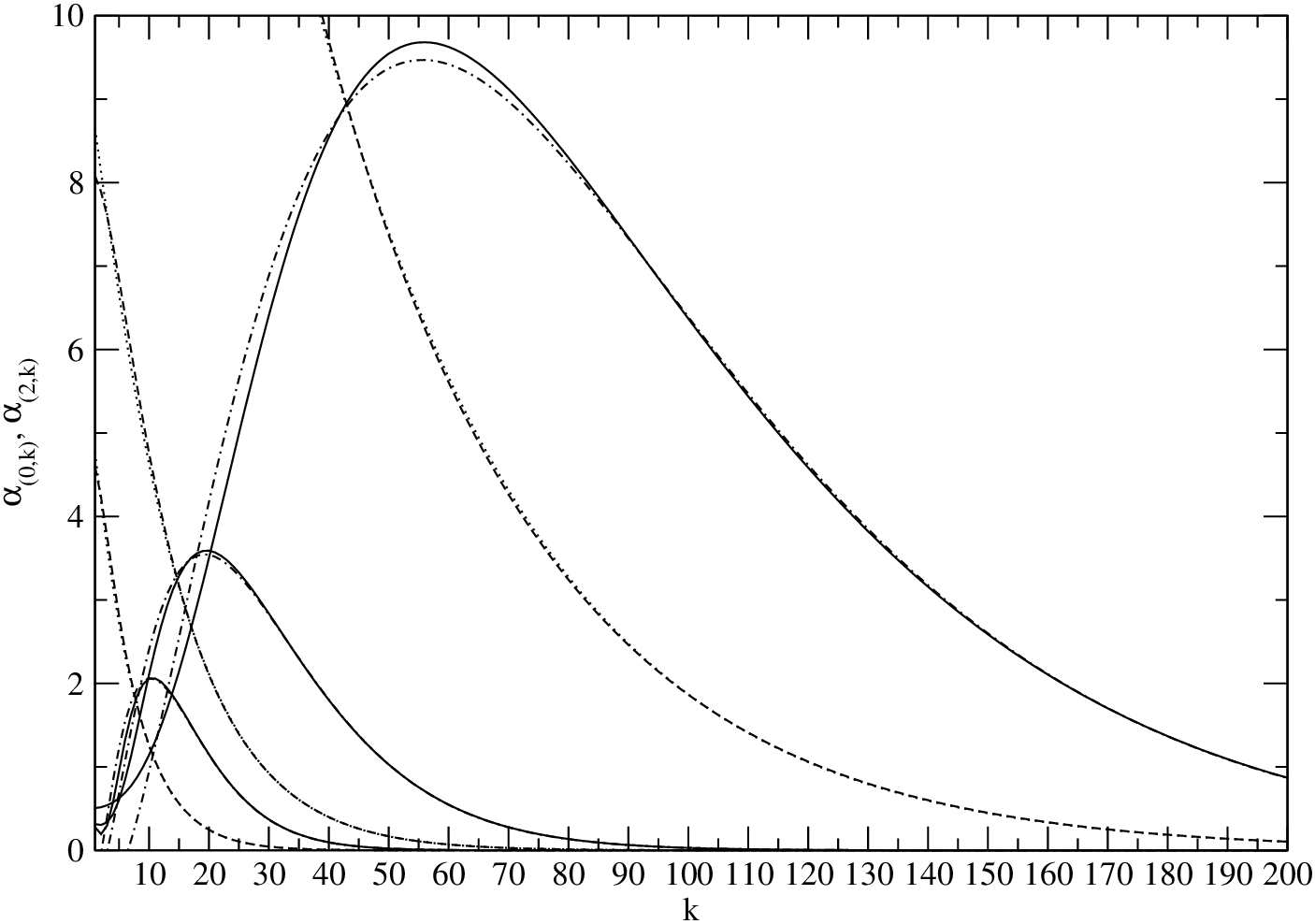}
\caption{ As in \ref{fig1} and  \ref{fig2}
but for $e=0.7$, $0.8$  and $0.9$.
Curves of the same type with
increasing values of $\alpha_{m,k}$
correspond to increasing values of $e.$ \label{fig3}}
\end{figure*}

Here we briefly  compare  the approximate values
for  the Fourier coefficients $\alpha_{2,k}$ and $\alpha_{0,k}$
obtained with  our analytic approximations  derived in the Appendix with
 the exact  coefficients obtained numerically.
  The various coefficients are plotted in figures \ref{fig1},\ref{fig2} 
and \ref{fig3}.
In figure \ref{fig1} we  compare values 
of the  Fourier coefficients obtained analytically
with those obtained numerically
for small $e=0.2$ and $e=0.3$. One sees that
$\alpha_{0,k}$ decreases  monotonically with $k$
and that the analytic and numerically obtained values
are in a good agreement for practically all values of $k$. 

The difference  is largest  and of order of $10$ percent
at $k=1$  and rapidly decreases thereafter with $k$. 
The coefficient $\alpha_{2,k}$ has a maximum  at $k_{max}=2$ for $e=0.2$
and at $k_{max}=3$ for $e=0.3.$
For values of $k > k_{max}$ the
analytic and numerically obtained values
differ by at most $10$ percent and approach each other
rapidly with increasing $k$. 
For values of $k < k_{max}$ the agreement between the 
analytic and numerically values is less good.

In figure \ref{fig2} the same comparison is given
for intermediate values of eccentricity $e=0.4$, $0.5$ and 
$0.6.$
In this case the agreement between numerical and analytic
curves is much better. The disagreement  clearly 
decreases with increasing $e$ and for  $e=0.6$   the curves for $\alpha_{0,k}$ 
differ by less than or of the order of
$\sim 1$ percent.
In the case of $\alpha_{2,k}$ 
the agreement is less good, however, in contrast with the 
case  of smaller $e,$ the disagreement remains small 
even for $k<k_{max}$ provided that $k$ is sufficiently large.
 For example 
 the curves corresponding to $e=0.6$  differ by less
than $\sim 10$ percent for $k>4$. 

In figure \ref{fig3}
 we show the comparison for $e=0.7$, $0.8$  and 
$0.9.$ The disagreement between analytic and numerically obtained values
is very small for all interesting values of $k$. 
The numerical and analytic curves corresponding to $\alpha_{0,k}$
practically coincide with each other.
The  numerical and analytic curves corresponding 
to $\alpha_{2,k}$ and $e=0.9$ differ  by less than $\sim 1$ percent
for $k>10$.  

In general the analytic
and numerical curves
are in better agreement   for $\alpha_{0,k}$
than for $\alpha_{2,k}$. 
This occurs because
in order to obtain our analytic estimate 
for $\alpha_{2,k}$ we had to expand 
$\cos(2\Phi)$ in powers of the eccentric anomaly $\xi$.
We expanded $\cos(2\Phi)$ to fourth order in $\xi$. 
It seems that one can obtain a better approximation 
by expanding  to  higher order in of $\xi$.

We point out that
our approximate analytic expressions for the Fourier
coefficients associated with 
the tidal amplitudes may be  useful in  a much more general situation
than the one considered here
 but where 
an analytic form of the tidal potential is required 
and a perturber moves on an eccentric orbit. 

Such situations could naturally 
arise, say, in the
problem of  the interaction of an accretion disc 
with a perturbing planet or star, or in a problem of  the interaction of 
a super-massive black hole binary with a stellar cluster.

\section{Orbital evolution} \label{ORBE}
The evolution equations for the orbital parameters follow from equations
(\ref{eqno 14}) and equations  (\ref{eqno 35}), (\ref{eqno 36})  
 giving  two equations for the rate of change of orbital
semi-major axis and  specific angular momentum in the form

\be \dot a =-{2\dot E_{\nu}^{0}a^{2}\over GMm}(\Psi_{1}-2\sigma \Psi_{2})=
-{3\pi\over 5}{(GM)^{2}Q^{2}\bar 
\gamma (1+q)\over M_{p}\omega_{0}^{4}}
{1\over a^{7}}(\Psi_{1}-2\sigma \Psi_{2}), \label{eqno48} \ee
\be \dot L_{orb}={\dot L_{\nu}^{0}\over m}(\Psi_{2}-2\sigma \Psi_{3})=
{3\pi\over 5}{(GM)^{5/2} Q^{2}\bar 
\gamma \sqrt{(1+q)}\over M_{p} \omega_{0}^{4}}{1\over a^{15/2}}
(\Psi_{2}-2\sigma \Psi_{3}), \label{eqno49}\ee
where $L_{orb}$ is the specific orbital angular momentum and we recall that $a$ is the semi-major axis,
$M_{p}$ is the mass of the primary, $q=M_{p}/M$.

To determine the ratio of the primary rotation rate and the orbital 
mean motion, $\sigma = \Omega_r /\Omega$, 
and hence
close the set of equations (\ref{eqno48}) and (\ref{eqno49}) 
we use the law of conservation of angular momentum in the form:

\be L_{orb}+I\Omega_{r}=L_{in}, \label{eqno 50}\ee
where $I$ is the moment of inertia of the primary per unit mass and $L_{in}$  
taken to be the 'initial' value of the 
specific orbital angular momentum 
at the beginning of  the  orbital evolution due to equilibrium tides. 
Note that in addition to $\sigma,$  the quantities 
$\Psi_{1}$, $\Psi_{2}$ and $\Psi_{3}$ 
depend on eccentricity $e$ which itself may be 
expressed in terms of $a$ and $L_{orb}$  by use of 
the standard relation $e=\sqrt{1-{L_{orb}^{2}\over GMa}}.$      

It is convenient to rewrite equations (\ref{eqno48}) 
and (\ref{eqno49}) in dimensionless form
by use of a length scale $,a_{0},$ which is   
defined to be the semi-major axis corresponding to the tidal equilibrium state 
of the dynamical system found from the conditions $( da/dt = dL_{orb}/dt =0).$ 

When $a=a_{0}$ is determined in this way we also have 
$e=0$, $\sigma=1$ in equilibrium. 
We define the dimensionless semi-major axis, $x$, to be 
$x=a/a_{0}$ and the dimensionless angular momentum  to be 
$y= L_{orb}/L_{0}$, where $L_{0}=\sqrt{{GMa_{0}\over 1+q}}$.   

In terms of these variables, equations 
(\ref{eqno48}) and (\ref{eqno49})  take the form
\be {dx\over d\tilde t} =-{1\over x^{7}}(\Psi_{1}-2\sigma \Psi_{2}), \label{eqno 51} \ee
\be {dy\over d\tilde t}={1\over x^{15/2}}(\Psi_{2}-2\sigma \Psi_{3}), \label{eqno 52}\ee
where we introduce   the  dimensionless time $\tilde t=t/t_{\nu}$  with 
\be t_{\nu}={5\over 3\pi}{M_{p}\omega_{0}^{4}a_{0}^{8}\over (GMQ)^{2}\bar \gamma (1+q)} 
\label{eqno 53}\ee
The time $t_{\nu}$ defines a characteristic time scale for tidal evolution
of the orbital elements, in this case  due to equilibrium tides.

We  note that in terms of the dimensionless variables, 
the eccentricity $e=\sqrt {(1-y^{2}/x)}$ and
the law of conservation of angular momentum (\ref{eqno 50}) takes the form
\be y+\tilde I{\sigma \over x^{3/2}}=y_{in}, \label{eqno 54}\ee
where $y_{in}=L_{in}/\sqrt{GMa_{0}}$. 
We recall that $\sigma=\Omega_{r}/\Omega$ and
a dimensionless moment of inertia
$\tilde I=I(1+q)/a_{0}^{2}$. 

As for most cases of astrophysical interest, the angular momentum content
of the orbit dominates that associated with the rotation of the primary, 
for simplicity,
we consider below only the case when the primary effectively  has low-inertia 
assuming  that $\tilde I \ll 1.$

As we see below, in this case,  the orbital evolution of the binary is  essentially
independent of the value of $\tilde I$.
 When the viscosity Fourier transform  coefficients $\tilde \gamma_{m,k}$ are specified, 
the set of equations (\ref{eqno 51}), (\ref{eqno 52}) 
and (\ref{eqno 54}) becomes complete.

We note  that neither the length scale $a_{0}$ or $q$ enter  
into   the above dimensionless set  of equations
or the  associated dimensionless initial conditions making the problem
scale free. 
That means that
solutions of the tidal problem can be scaled 
in a self-similar manner through  adjusting this variable. Note that in 
order to make such a similarity
transformation in general, the parameters entering into  the 
specification of  the viscosity coefficient 
 may also  need to be scaled appropriately.

\subsection* {Explicit form of the viscosity coefficients}
Before solving the equations of tidal evolution 
we should specify the viscosity Fourier transform  coefficients $\gamma_{m,k}.$
As we have mentioned above, it was suggested by Goldreich  \& Nicholson (1977)
and Goldreich \&  Keeley (1977) that  the action 
of an effective viscosity  arising from convective turbulence
should  be suppressed 
when the typical time scale  associated with convective eddies is much larger
than the time scale associated with the global disturbance
on which it is presumed to act.
To take this effect into account 
in the simplest  manner possible, we  adopt expressions for $\gamma_{m,k}$ of the form
\be \gamma_{m,k}={\gamma_{0}\over (1+(\omega_{m,k}t_{c})^{p})}, \label{eqno 55}\ee
where
\be \omega_{m,k}=|k\Omega +m\Omega_{r}| \label{eqno 56}\ee
is   the apparent frequency  associated with  the global disturbance as 
viewed  in the frame co-rotating with the primary, it being assumed that that
is well defined.
\footnote{We stress that in the case of forced oscillations,
the responding  global disturbance with particular $m$ and $k$ oscillates with the 
frequency of the corresponding Fourier term in the expansion of the tidal forcing amplitude.}
Here $p$ is a constant parameter. 

Goldreich  \& Nicholson  (1977) and Goldreich \& Keeley (1977) adopted  $p=2.$
However, Zahn(1989) has  argued that $p\sim 1.$  Here 
we shall allow for general values of $p$. 
Equation (\ref{eqno 55}) can be
brought into the standard form (\ref{eqno 22}) with
\be \tilde \gamma_{m,k}={T_{*}^{p}\over (1+T_{*}^{p}|k+m\sigma|^{p}x^{-3p/2})},
\quad \bar \gamma= {\gamma_{0}\over T_{*}^{p}}, \label{eqno 57}\ee
where the dimensionless parameter $T_{*}=t_{c}\sqrt{GM\over a_{0}^{3}}$ 
determines the magnitude of the suppression of  the viscosity. 
We assume below that $T_{*} \gg 1$. In this limit, many details of the orbital evolution are 
essentially independent of the particular value of this parameter.

\subsection*{ Effective spin orbit  resonances}
A feature arising  in the limit $T_{*} \gg 1$ is that the action of viscosity
is most effective  for a particular forcing term
when the rotational and orbital frequencies are such that $(k+m\sigma)=0.$
When this satisfied the potential  pattern associated with the Fourier component
 $(m,k)$  of the forcing
potential co-rotates with the primary. The relative forcing frequency is thus zero
and there is no suppression of the viscosity.
When $T_{*} \gg 1$ viscosity acts strongly only very close to the corotation
resonances with $(k+m\sigma)=0,$ occurring when  $(k=1,2,3...., m=-2).$ 
Hence,  the terminology associated with a  resonance is used.

As we shall see below, the qualitative nature 
of the orbital evolution  depends on  whether the parameter $p$ exceeds or is less
than unity. This is related to the resonant behaviour considered above. 

Let us suppose that $T_{*} \gg 1$ and we are close to a resonance such that the value of 
the dimensionless rotational frequency $\sigma$ is close to some integer, $n,$
divided by two, thus  $\sigma \approx n/2$. 
Then it
follows from equation (\ref{eqno 30c}) and (\ref{eqno 31c}) that the quantities
$S_{1}=(\Psi_{1}-2\sigma \Psi_{2})$ and  
$S_{2}=(\Psi_{2}-2\sigma \Psi_{3})$ 
are mainly determined by the terms in the respective summations for these expressions
with $k=n$ and $m=-2$ which are resonant in the above sense. 
Later we  call these terms the 'resonant terms'.

We have $S_{1}=S_{1}^{r}+S_{1}^{nr}$ and $S_{2}=S_{2}^{r}+S_{2}^{nr}$, where   
\be S_{1}^{r}={2T_{*}^{p}n(n-2\sigma)\phi_{2,-n}^{2}\over (1+T_{*}^{p}|n-2\sigma|^{p}x^{-3p/2})}, \quad 
S_{2}^{r}={2T_{*}^{p}(n-2\sigma)\phi_{2,-n}^{2}\over (1+T_{*}^{p}|n-2\sigma|^{p}x^{-3p/2})}, \label{eqno 58}\ee
and $S_{1}^{nr}$, $S_{2}^{nr}$ stand for all other (non-resonant) terms in the series.

When we have the exact equality,
$\sigma=n/2,$ the resonant terms are equal to zero.
However, their absolute values increase very sharply with an  increase of
the difference $|\sigma-n/2|$.
For example, when $T_{*} \gg 1$ and $|\sigma - n/2|\sim T_{*}^{-1} \ll 1$, 
the resonant terms are proportional to $T_{*}^{(p-1)}$ in absolute magnitude.
Therefore, for the case $p > 1$ they can, in principal, be much larger than the non-resonant terms. 
On the other hand, in the case of $p< 1$ 
the contribution of the resonant terms to $S_{1}$ and $S_{2}$ is insignificant. We shall see below that in the case
$p > 1,$ the orbital evolution can proceed through a stage   
where the condition $\sigma \sim n/2$ is maintained for 
a long time. 
Thus, the system can evolve in a state that maintains 
a specific spin-orbit corotation resonance.

\subsection*{The case $p < 1$}

When the parameter $p$ is smaller than $1$, 
the orbital evolution of a low inertia primary star is,
in many details, similar to
the case of assumed
constant time lag between response and forcing potential
considered by Alexander (1973) and Hut (1981). This also
corresponds to the standard Navier Stokes viscosity $(t_c =0)$.
If the initial value of $\sigma$ is sufficiently small, because
of the small inertia of the primary, the system quickly
relaxes to a state where the orbital 
angular momentum is approximately conserved  
and we can assume that $y\approx y_{in}\approx 1$.
In this case 
the eccentricity and the semi-major axis are related to each other as
\be e\approx \sqrt{(1-1/x)}.\label{eqno 59}\ee 
Also because of the low inertia of the primary,
its rotation rapidly approaches the value given by  $\Omega_{r} \approx \sigma_{ps}\Omega$,
where $\sigma_{ps}$ corresponds to a state of so-called 
'pseudo-synchronisation' (Hut 1981) 
and is obtained from equation (\ref{eqno 52}) after setting 
$dy /d\tilde t =0.$

In a state of pseudo-synchronisation, $\sigma_{ps}$   is given by 
\be \sigma_{ps}={\Psi_{2}\over 2\Psi_{3}}. \label{eqno 60}\ee
However, note that both $\Psi_{2}$ and $\Psi_{3}$ are,
in general, functions of $\sigma_{ps}$,
and therefore equation (\ref{eqno 60}) 
should
be considered as a non-linear algebraic equation for $\sigma_{ps}.$ 
It is convenient to express its solution in the form
$\sigma_{ps}=\sigma_{1}\sigma_{H}$ where $\sigma_{H}$
represents the solution of (\ref{eqno 60}) for the case of constant  time lag
or the standard Navier Stokes viscosity $(t_c=0).$ 
This is given by (Hut,1981)
\be \sigma_{H}={\Psi^{0}_{2}\over 2\Psi^{0}_{3}}=
{1\over \epsilon^{3}}{(1+{15\over 2}e^{2}+{45\over 8}e^{4}+{5\over 16}e^{6})\over (1+3e^{2}+{3\over 8}e^{4})}, \label{eqno 61}\ee  
$\epsilon=\sqrt{(1-e^{2})}$ and we have made  use  of equations
(\ref{eqno 46}) and (\ref{eqno 47}).

\begin{figure}
\vspace{8cm}\includegraphics{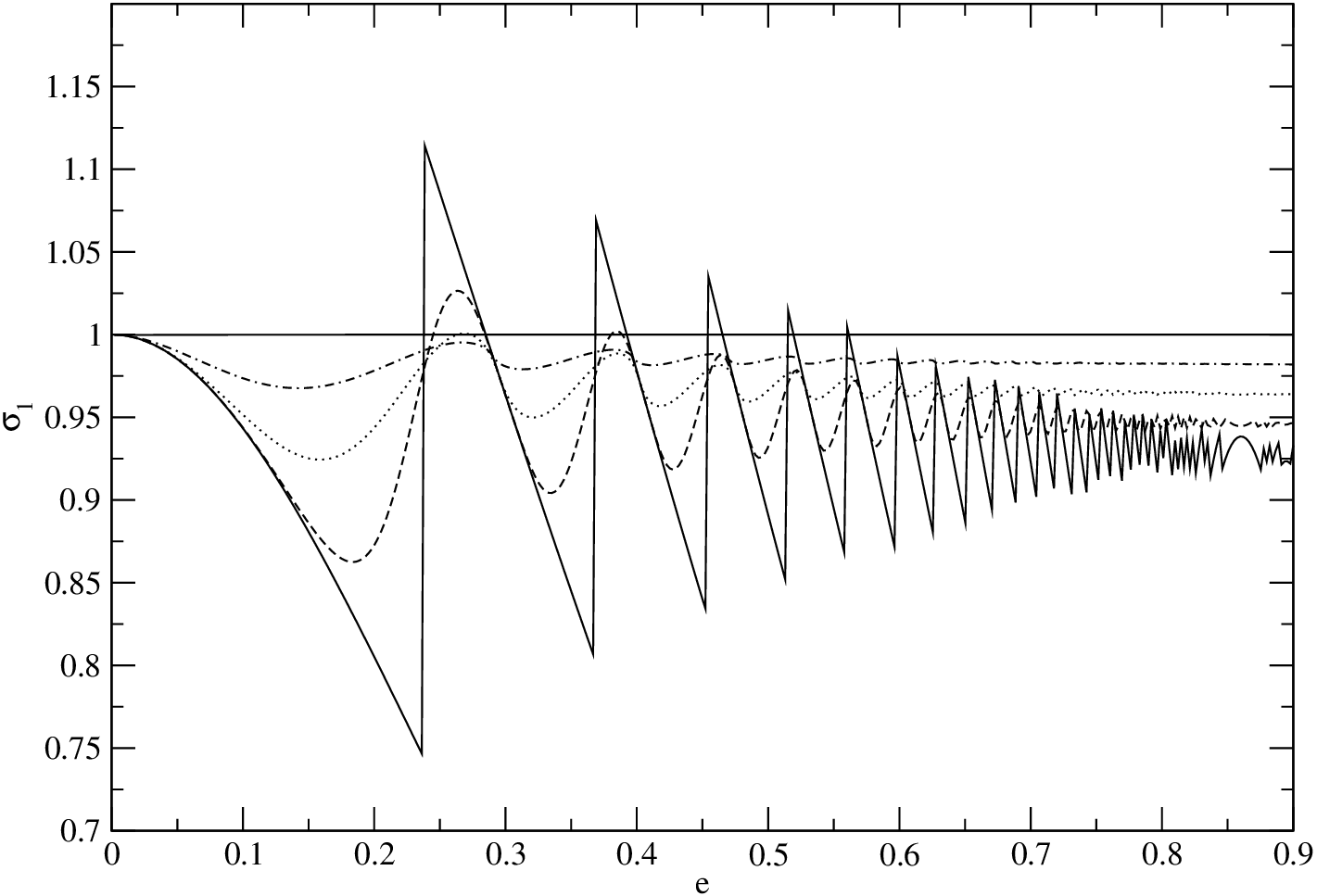}
\caption{The quantity $\sigma_{1}=\sigma_{ps}/\sigma_{H}$ as a function of $e$. The line $\sigma_{1}=1$ corresponds to $p=0$,
the solid, dashed, dotted and dot-dashed curves correspond to $p=1$, $0.75$, $0.5$, $0.25$, respectively. Note that the curve 
with $p=1$ exhibits an irregular behaviour at large eccentricities $e > 0.85$.
This is most likely to be a numerical artifact related to the
technical  problem of finding roots of equation
(\ref{eqno 60}) in the case of a very rapidly varying function. \label{fig4}}
\end{figure}

We show the dependence of $\sigma_{1}$ on $e$ obtained 
numerically for different values of $p$
in figure \ref{fig4}. We see that the difference between 
$\sigma_{ps}$ and $\sigma_{H}$ is rather small for
sufficiently large eccentricities. 
This comes about, in simple terms, because 
the inverse stellar rotation frequency $\Omega_{ps}^{-1}=1/(\sigma_{ps}\Omega)$
at pseudo-synchronisation 
is always of the order of a characteristic time of periastron passage. 

To obtain an approximate equation for the 
evolution of the semi-major axis we substitute equation
(\ref{eqno 60}) into  equation (\ref{eqno 51}) 
obtaining
\be {dx\over d\tilde t}={1\over x^{7}}(\Psi_{1}-{\Psi_{2}^{2}\over \Psi_{3}}). \label{eqno 62}\ee

As we have mentioned above, resonant effects are not important when 
$p< 1.$ That means that to a good approximation we
may neglect unity in the  denominator of the 
expressions ({\ref {eqno 57}) in the limit $T_{*} \rightarrow \infty$. 
Then the functions 
$\Psi_{1}$, $\Psi_{2}$ and $\Psi_{3}$ in equation 
(\ref{eqno 62}) do not depend on the parameter $T_{*}$. Taking into account 
equation
(\ref{eqno 59}) we see that for a given $p$ 
the right hand side of equation (\ref{eqno 62}) is only a
function of the semi-major axis $x$. In that case, the variables in equation
(\ref{eqno 62})
can be separated and  equation (\ref{eqno 62}) can easily 
integrated by numerical means.
We show the result of integration of equation
(\ref{eqno 62})
in figures \ref{fig5} and \ref{fig6} together with the results of 
numerical integration of the full set of equations (\ref{eqno 51}), (\ref{eqno 52}) and (\ref{eqno 54}).
One can see that the approximate semi-analytic approach gives a very good approximation 
to the solutions of the full set of equations.      

Note that for our numerical integration we set $e=0.9$ initially
and terminated the integration when
$e=0.2$. We use a very large value of $T_{*}=10^{5}$ and a small value of the dimensionless moment of
inertia $\tilde I=10^{-3}$. We have checked that other reasonable values of these parameters dot not lead
to a different evolution of our dynamical system.
We have used our approximate expressions for the Fourier coefficients 
of the tidal forcing amplitude in 
numerical integration of equations (\ref{eqno 51}), (\ref{eqno 52}) and (\ref{eqno 54}) and  good agreement
between numerical and semi-analytic calculations 
implies that use of the approximate analytic  expressions for the Fourier coefficients 
is justified.
 
\begin{figure}
\vspace{8cm}\includegraphics{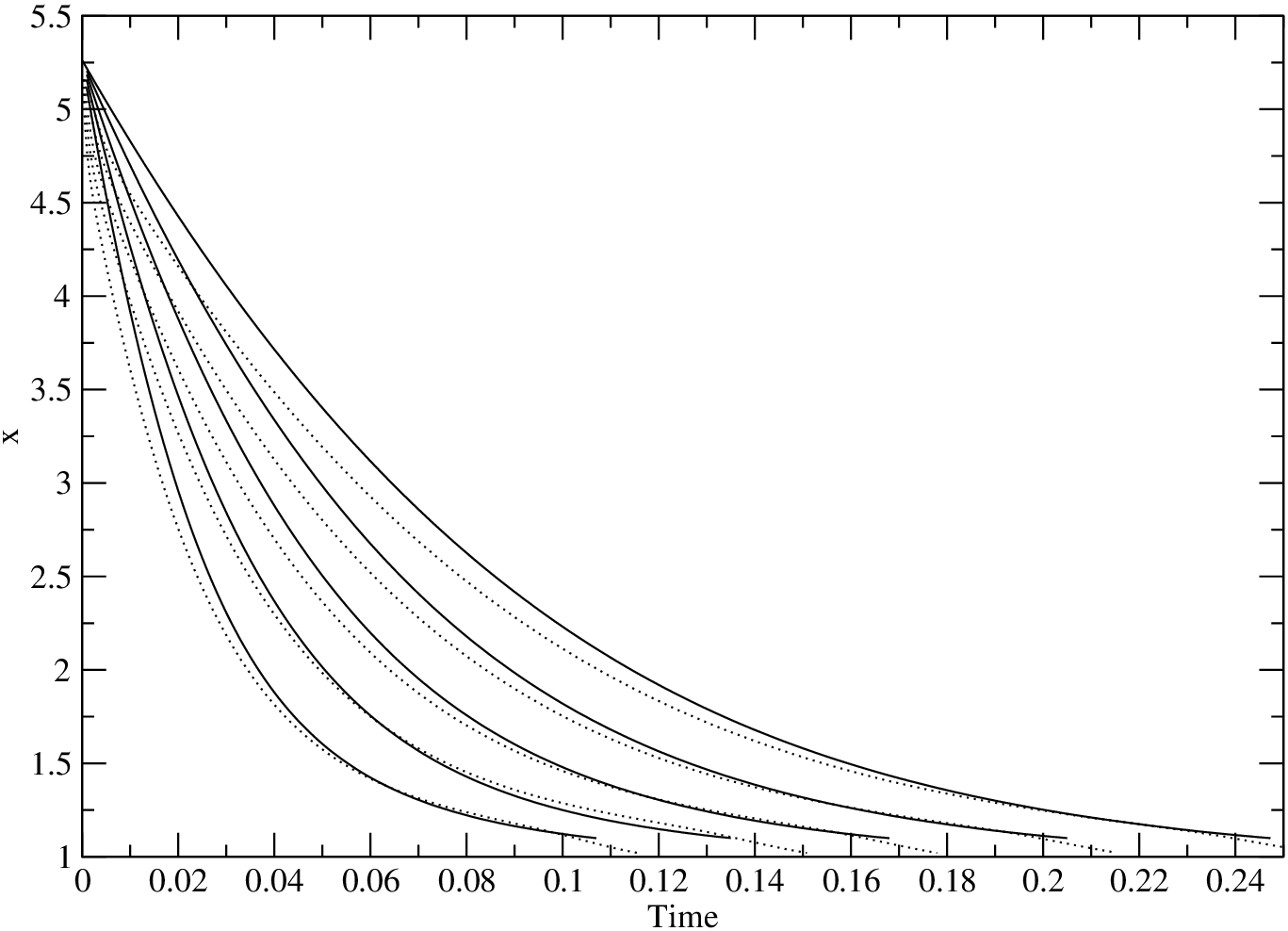}
\caption{The dependence of $x=a(t)/a_{0}$ on  dimensionless 
time  $\tilde t$. The solid curves are the solutions of equation (\ref{eqno 62}) and
the  dotted curves
represent the solutions of the full dynamical system.
Different curves of the same type correspond to  $p=0$, $0.25$,  
$0.5$, $0.75$ and $1$.  Curves having larger values of $x$ at the same time $\tilde t$  
have larger values of $p$. The initial value of $\sigma$ is set to zero}
\label{fig5}\end{figure}

\begin{figure}
\vspace{8cm}\includegraphics{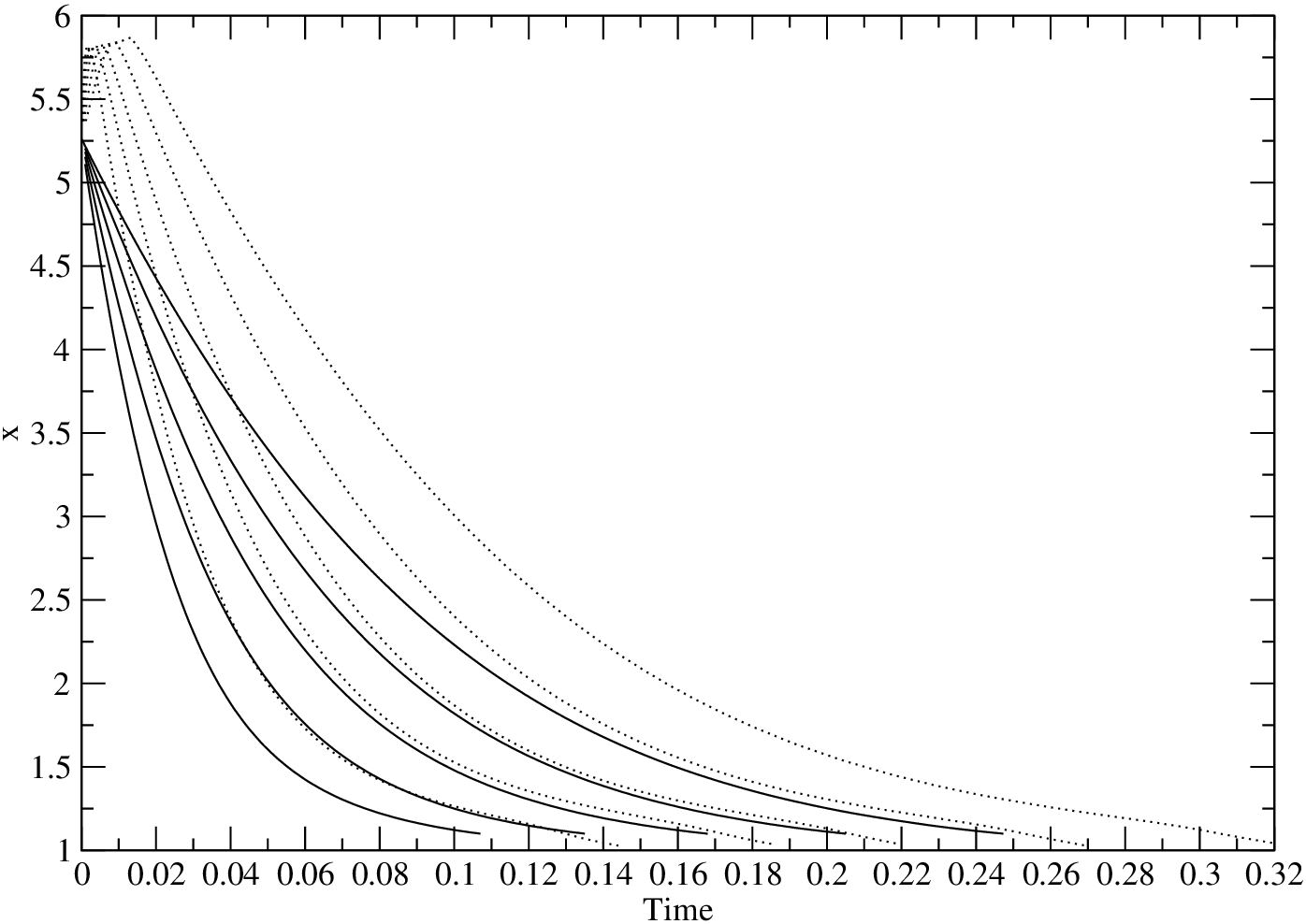}
\caption{Same as figure \ref{fig5},
but the initial value of $\sigma$ is set to $100$.}
\label{fig6}\end{figure}

In figure \ref{fig5}
we show numerical solutions of our equations setting initial value of the rotational parameter
$\sigma_{in}$ equal to zero,
and in figure 6 we consider the case of a high $\sigma_{in}=100$.

From these and similar plots the dimensionless time required to approach circularity
starting from large eccentricity can be estimated.
This is typically $\sim 0.2.$
Both cases show
quite similar late time evolution.
However, in the case of $\sigma_{in}=100$ we also get a short initial period 
of time when the value of semi-major axis increases with time.
During this period of time the rotational
energy of the primary is transfered to the orbital energy.
We discuss this effect in more detail later on. 

\subsection{The case $p > 1$}

In this case the evolution is qualitatively different from
when $p < 1.$
This is because of the effects of spin orbit resonances become
important.
To see this 
let us consider the case $p=2$ 
where a simple analytic approach is possible in detail.

\subsubsection{pseudo-synchronisation}
Assume that the moment of inertia is very small and the  rotation of the
primary is close to a
spin-orbit resonance of the order $n$ so that 
$\Delta \sigma =\sigma -n/2 \ll 1.$ 
Then just as for the case $p < 1$ the assumption of low primary inertia
means we can consider
the orbital angular momentum to be
a conserved quantity and set $ dy/d\tilde t = 0.$ 

It then follows from equation (\ref{eqno 52}) that
$S_{2}=\Psi_{2}-2\sigma \Psi_{3}=0.$ 
As discussed above we divide $S_{2}$ into resonant and non-resonant parts
writing $S_{2}=S_{2}^{r}+S_{2}^{nr}$ 
 we must then have $S_{2}^{r}=-S_{2}^{nr}.$ 
Using the explicit form
of $S_{2}^{r}$  (see equation  (\ref{eqno 58})) we obtain
\be {\Delta \sigma \over 1+4T_{*}^{2}x^{-3/2}(\Delta \sigma)^{2}}
={S_{2}^{nr}\over 4T_{*}^{2}\phi_{2,-n}^{2}}. \label{eqno 63}\ee

The solution of this equation gives the
deviation of $\sigma$ from the value corresponding to exact resonance as
\be \Delta \sigma ={x^{3}\phi_{2,-n}^2\over 2S_{2}^{nr}}
\left(1 -
\sqrt{1-{(S_{2}^{nr})^{2}\over x^{3}T_{*}^2\phi_{2,-n}^{4}}}\right ),
\label{eqno 64}\ee 
where we do not consider the second unphysical root. 
For a physically meaningful solution
the expression contained within the square root  in
equation (\ref{eqno 64}) must be positive.
This implies that the dynamical system can
stay in the resonance only if
\be {|S_{2}^{nr}|\over \phi_{2,-n}^{2}}\le x^{3/2}T_{*}, \label{eqno 65}\ee
and therefore
\be |\Delta \sigma | \le \Delta \sigma_{max}={x^{3/2}\over T_{*}}. \label{eqno 66}\ee
Condition (\ref{eqno 65}) tells that for a given $T_{*}$ and $x$,
evolution of the dynamical system is possible only
if the ratio ${|S_{2}^{nr}|\over \phi_{2,-n}^{2}}$ is sufficiently small. 
This can always be satisfied if $T_*$ is sufficiently large.

However, for a sufficiently large $n$ this ratio
increases with decrease of eccentricity $e$. Therefore, 
during the process of tidal circularization, a resonance 
with some particular order $n=2\sigma$ becomes at some point impossible.
The dynamical system may then 
rapidly relax to some other resonance 
with $n' < n$ where the ratio is smaller and
the condition ($\ref{eqno 65}$) is satisfied.
Thus, the analytic approach indicates that a low
inertia primary evolves through a sequence of spin-orbit resonances with 
decreasing  $n$ during the process of tidal
circularization. 
As we see below this character 
of the  evolution is confirmed by numerical simulation. Note that the
qualitative character of the tidal evolution remains the same for any $p > 1$. 

\subsubsection{Evolution of the semi-major axis}

Now let us discuss the evolution of the semi-major axis.
We assume that condition ($\ref{eqno 65}$) is 
satisfied and the primary evolves in a state of spin-orbit resonance 
with some particular $n=2\sigma$. In order to find 
an evolution law for the semi-major axis we have to evaluate $S_{1}=\Psi_{1}-2\sigma \Psi_{2}$. In general, this
quantity can be represented only in terms of complicated 
summations  (see  equation \ref{eqno 30c}). However, for the case
$p=2$ these series can be  performed in the limit
$T_{*}\rightarrow \infty $. 
The resulting expression 
is remarkably
simple.
Then we can go on to  obtain an analytic expression 
for the dependence $x(\tilde t)$ (see below).

The quantity $S_{1}$ can also be divided on the resonant and non-resonant parts (see equation \ref{eqno 58}): 
$S_{1}=S_{1}^{r}+S_{1}^{nr}$. It is easy to see that $S_{1}^{r}=nS_{2}^{r}$ and from the condition of
constant orbital angular momentum it follows that $S_{2}^{r}=-S_{2}^{nr}$, and we have
\be S_{1}=S_{1}^{nr}-nS_{2}^{nr}.\label{eqno 67}\ee
 Now the explicit form of $S_{1}$ can be  written down
with help of equations (\ref{eqno 30c}) and (\ref{eqno 31c}) in the form

\be S_{1}=2\sum_{k\ne n}(k-n)(k-2\sigma)\tilde \gamma_{2,-k}\phi^{2}_{2,-k}+    
(k+n)(k+2\sigma)\tilde \gamma_{2,k}\phi^{2}_{2,k}+
{2\over 3}k^{2}\tilde \gamma_{0,k}\phi_{0,k}^{2}. \label{eqno 68}\ee
In equation (\ref{eqno 68}) we can set $2\sigma = n.$
We can also neglect unity in the denominators of the expressions
for $\tilde \gamma_{m,k}$ and set $2\sigma = n$ there 
(see equations (\ref{eqno 57})) so as to obtain
\be \tilde \gamma_{2,\mp k}={x^{3}\over (k\mp n)^2}, \quad
\tilde \gamma_{0,k}={x^{3}\over k^2}, \label{eqno 69}\ee

Using equation (\ref{eqno 69}) in equation (\ref{eqno 68}) 
gives 
\be S_{1}=2x^{3}\sum_{k=1}^{\infty} (\phi_{2,-k}^{2}+
\phi_{2,k}^{2}+{2\over 3}\phi_{0,k}^{2})=
x^{3}\sum_{k=1}^{\infty} (\alpha_{2,k}^2+
\beta_{2,k}^{2}+{1\over 3}\alpha_{0,k}^{2}), \label{eqno 70}\ee
where in the last equality we use expressions of $\phi_{m,k}$ 
in terms of $\alpha_{m,k}$ and $\beta_{m,k}$
(see equation (\ref{eqno 19})).
Note that the series (\ref{eqno 70}) does not contain any divergence 
at $k=n$ and we include the term 
with $k=n$ in the series, assuming with justification
when $n$ is large, that it gives a negligible contribution. 

According to the Parseval's theorem we have
\be \sum_{k=1}^{\infty} (\alpha_{2,k}^2+\beta_{2,k}^{2})=
{1\over \pi}\int_{0}^{2\pi}d\tau (\phi_{+}^{2}+\phi_{-}^{2}),
\quad 2\alpha_{0,0}^2
+\sum_{k=1}^{\infty} \alpha_{0,k}^2
={1\over \pi}\int_{0}^{2\pi}d\tau \phi_{0}^{2}. \label{eqno 71}\ee
Here we recall that the coefficients $\alpha_{2,0}$ and $\beta_{2,0}$ vanish 
and the
inclusion of these terms in the summations is redundant.
Furthermore the effect of $\alpha_{0,0}$ in the second 
summation  is negligible for orbits of significant eccentricity.
From definitions of the quantities $\phi_{\pm}$ and $\phi_{0}$ 
(see equation (\ref{eqno 21})) we have 
$\phi_{+}^{2}+\phi_{-}^{2}=\phi_{0}^{2}$.
Accordingly, both integrals in equation (\ref{eqno 71})   may be taken to be  
$\Psi_{3}^{0}={2\over \epsilon^{9}}(1+3e^{2}+{3\over 8}e^{4})$ 
(see equation (\ref{eqno 47})) with
$\epsilon=\sqrt{(1-e^{2})}$, and
\be S_{1}={4\over 3}x^{3}\Psi_{3}^{0}=
{8\over 3}{x^{3}\over \epsilon^{9}}\left(1+3e^{2}+
{3\over 8}e^{4}\right). \label{eqno 71a}\ee
Taking into account that in our case the orbital angular momentum is
approximately conserved we can express $e$ in terms of $x$ using 
equation (\ref{eqno 59}) and substituting
equation
(\ref{eqno 71a}) in equation (\ref{eqno 51}) we obtain a remarkably simple equation for evolution of
the semi-major axis  which takes the form
\be {dx\over d\tilde t}=-{35\over 3}x^{-3/2}\left(x^{2}-{6\over 7}x+
{3\over 35}\right). \label{eqno 72}\ee
Note that equation (\ref{eqno 72}) can be easily generalised to take into account the term $\alpha_{0,0}$ and
also the terms determined by a non-zero rotational angular momentum of the primary. 
The solution of equation  
(\ref{eqno 72}) can be expressed in terms of elementary functions
\be \tilde t=-\left ({6\over 35}\sqrt{x}+{1\over 10}
\sqrt{{15\over 2}}\ln{(z)}\right)+\tilde C, \label{eqno 73}\ee
where $\tilde C$ is a constant of integration and
\be z= \left|{\sqrt x -\sqrt x_{+}\over \sqrt x +\sqrt x_{+}}\right|^{c_{+}}
\left|{\sqrt x +\sqrt x_{-}\over 
\sqrt x -\sqrt x_{-}}\right|^{c_{-}} \label{eqno 74},\ee
where $x_{\pm}={3\over 7}(1\pm 2\sqrt{ {2\over 15}})$ 
and $c_{\pm}=x_{\pm}^{3/2}/2$.
It is very
important to note that solution (\ref{eqno 73}) does not depend
on the order $n$ of some particular resonance.
Therefore it describes approximately  
not only the 
evolution of the dynamical system in a state of spin-orbit resonance with
 particular  value of $n$,
but also a system evolving from one resonance to another but spending most of
its time in a state of spin-orbit resonances with different values of $n$.

\subsection*{Numerical calculations}

\begin{figure}
\vspace{8cm}\includegraphics{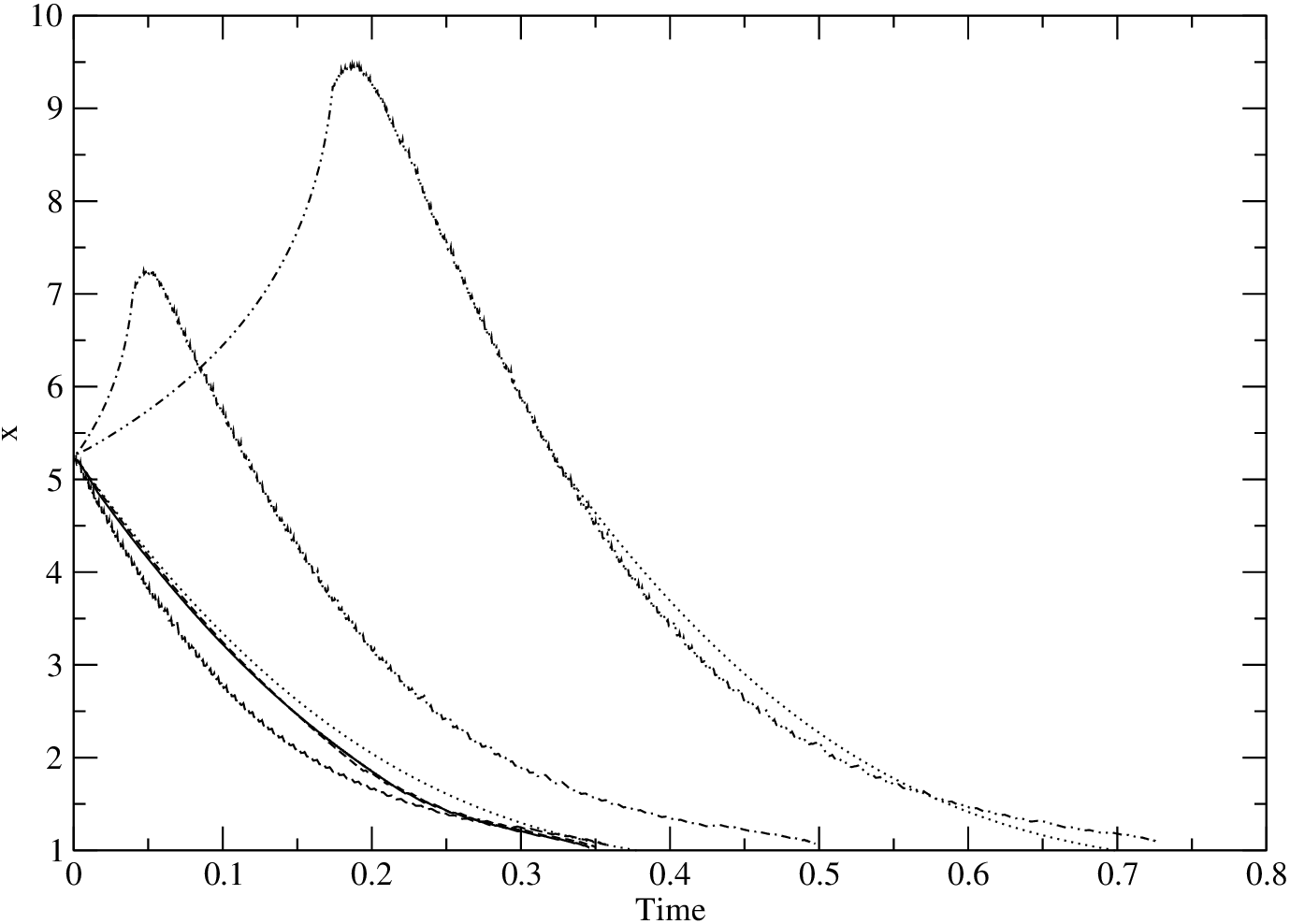}
\caption{The dependence of $x=a/a_{0}$ on time $\tilde t$. Different curves correspond to 
different values of 'initial' rotational parameter $\sigma_{in}$
set in the beginning of computations. The solid, short-dashed,
long-dashed, dot-dashed, and dot-dot-dashed curves correspond 
to $\sigma_{in}=0$, $10$, $50$, $75$ and $100$, respectively.
The dotted curves represent the analytic solution (\ref{eqno 73}) with different values of the integration constant $\tilde C$.
Note that the curves corresponding to $\sigma_{in}=0$ and $10$ (the solid and short-dashed curves, respectively), 
practically coincide.}  \label{fig7}
\end{figure}

We solve equations
 (\ref{eqno 51}) and (\ref{eqno 52}) 
numerically for different initial values of the rotational parameter 
$\sigma_{in}$ and $p=2$, and also 
for $\sigma_{in}=0$ and $p=1.5$ and $p=2.5$. 
The dimensionless parameter $T_{*}$ is set to $500$ and 
the dimensionless moment of 
inertia $\tilde I$ is set to $1/300$ for all numerical calculations. We have checked that other values of these parameters
do not change the dynamical evolution significantly provided that $T_{*}$ is sufficiently large and $\tilde I$ is sufficiently
small. All calculations are started with $e=0.9$ and terminated when $e=0.2$. 

 We start by discussing
 the case $p=2$. In figure \ref{fig7} we show the dependence of $x=a/a_{0}$ on time  $\tilde t$.
It is seen from this figure that the analytic solution (\ref{eqno 73}) is in a very good agreement with the numerical
solutions corresponding to $\sigma_{in}=0$ and $10$. It also approximates rather closely the case of 
intermediate $\sigma_{in}=50$. Results of calculation with high initial values of $\sigma=75$ and $100$ are in a good
agreement with our analytic expression only for sufficiently large values of time $\tilde t$. In the case of a large
$\sigma_{in}$, initially,
the semi-major axis (and eccentricity) grows with time
\footnote{Note that for all cases considered hereafter the orbital angular momentum is conserved in a good approximation
and equation (\ref{eqno 59}) is valid.}. Accordingly, the orbital energy also increases in the beginning of evolution. It turns out
that the rotational energy of the primary is transfered to the orbital energy on a short time scale and this drives an 
increase of the semi-major axis.     
\begin{figure}
\vspace{8cm}\includegraphics{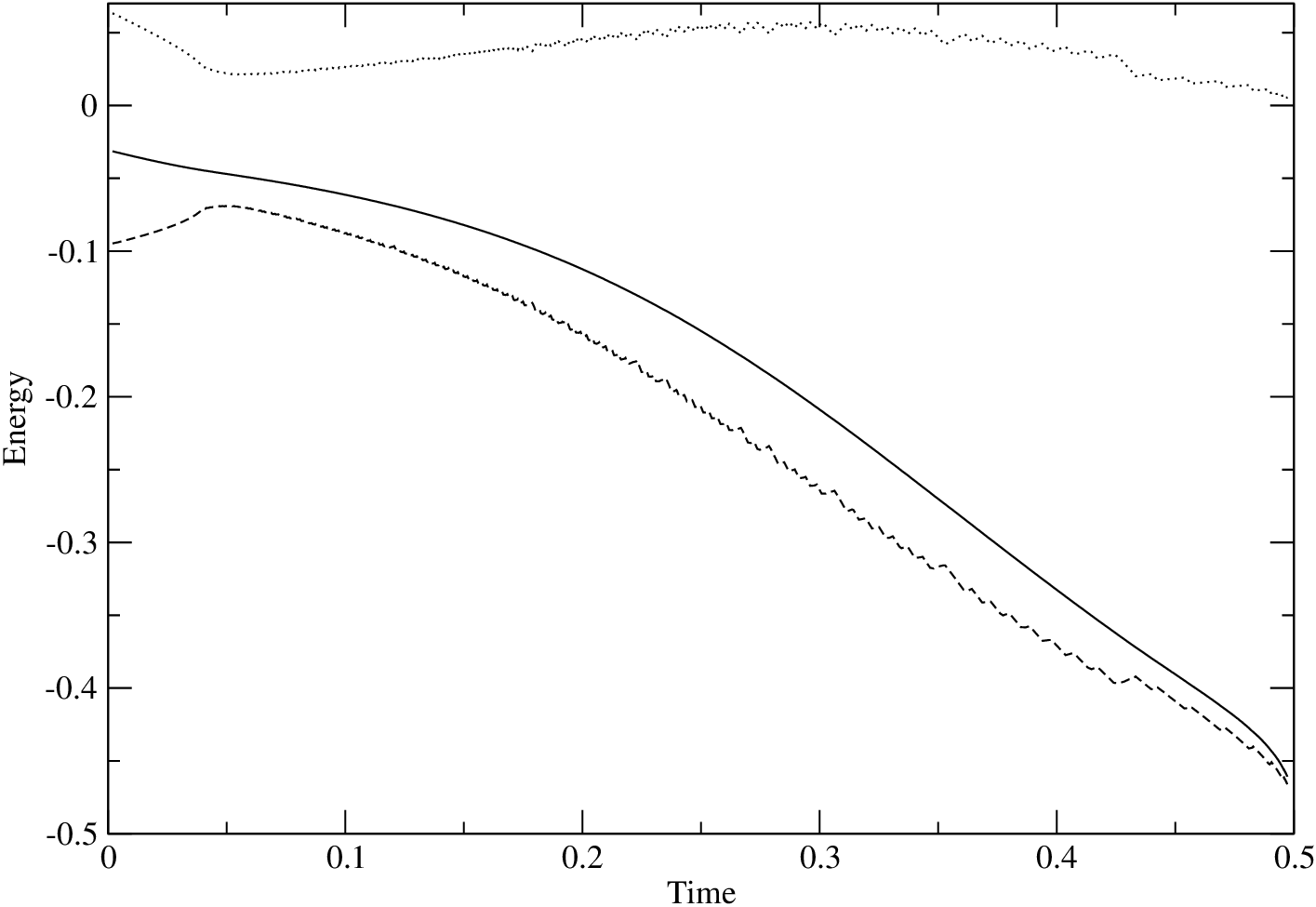}
\caption{The dependencies of the orbital energy (the dashed curve), the rotational energy 
(the dotted curve) and the sum of the orbital and rotational energies (the solid curve)
on time  $\sigma_{in}$.  All quantities are expressed in units of $GMM_{p}/a_{0}$. $\sigma_{in}$ is set to $75$. \label{fig8}}
\end{figure}
In figure \ref{fig8} we show the evolution of the orbital and rotational energies per unit of mass and their sum with time
for the case $\sigma_{in}=75$. The sum is gradually decreasing with time. That means that is it dissipated by the 
frictional processes. However, the orbital energy sharply increases and the rotational energy sharply decreases at
time $\tilde t < \tilde t_{1}\approx 0.05$. As it is seen from figure 7 the moment of time $\tilde t_{1}$ corresponds to
the maximum of the curve $x(\tilde t)$ with $\sigma_{in}=75$.  
\begin{figure}
\vspace{8cm}\includegraphics{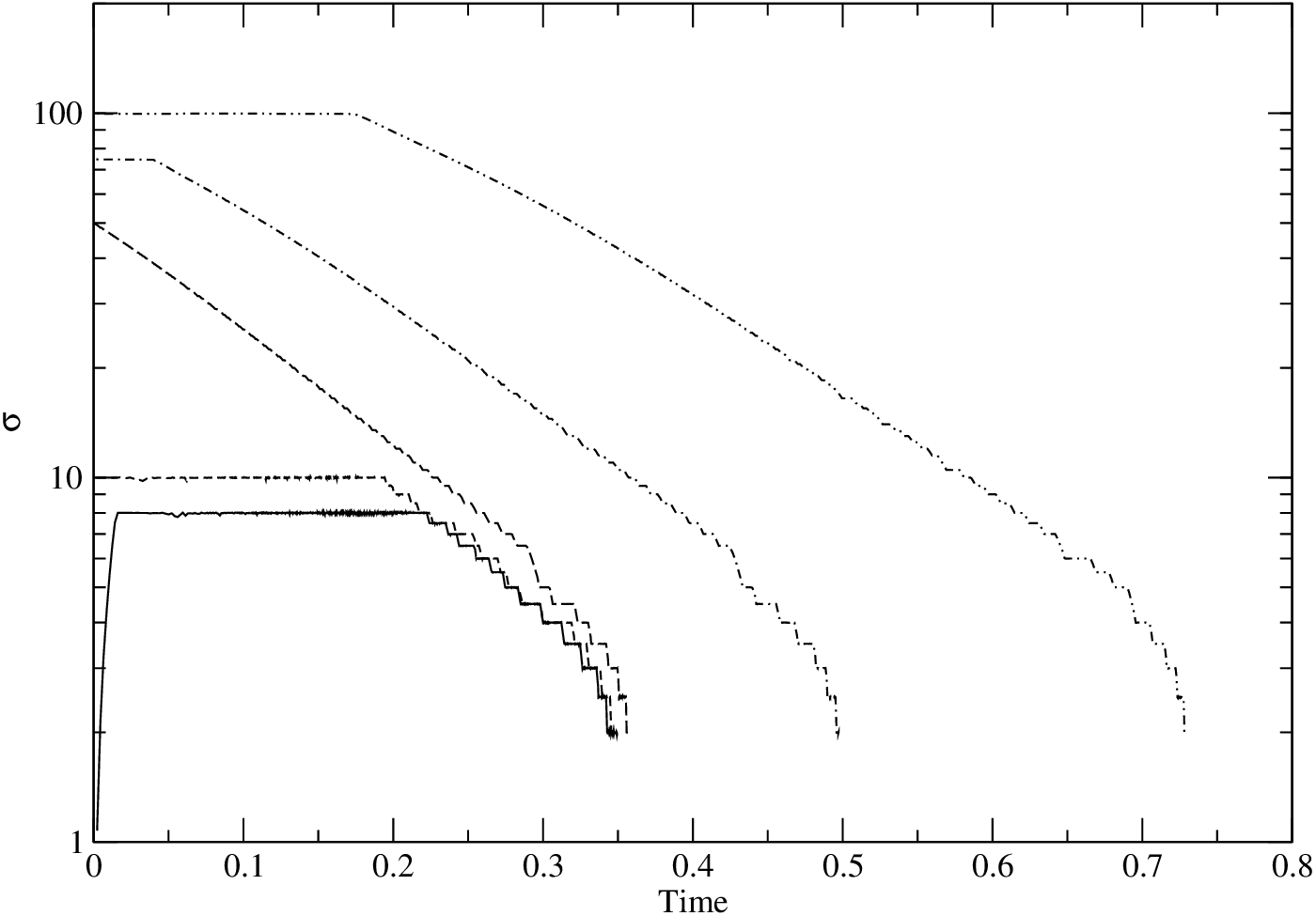}
\caption{The same as Fig. 7 but the dependence 
of $\sigma$ on time $\tilde t$ is shown.} \label{fig9}
\end{figure}
In figure \ref{fig9} we show the dependence of $\sigma$ on time for the cases with different $\sigma_{in}$. The solid curve corresponding to
$\sigma_{in}=0$ shows a sharp growth of $\sigma$ with time to the value $\sigma=8$. Then the dynamical system is evolving in the
$\sigma=8$ resonance for a
period of time $\tilde t < \tilde t_{2}\approx 0.23$. After the time $\tilde t_{2}$ the system is evolving through a sequence of
resonances with decreasing values of $\sigma$. The solution corresponding to $\sigma_{in}=10$ is evolving in the $\sigma=10$ resonance
from the very beginning. The late time evolution is similar to the previous case. The initial evolution of $\sigma$ for the case
$\sigma_{in}=50$ looks like a monotonic decrease of $\sigma$. In fact, as we see in the next figure 10 the system is evolving through
a large number of resonances with decreasing resonance order $n=2\sigma$, but does  not stay in any of these resonances for a
long time. The curves corresponding to $\sigma_{in}=75$ 
and $100$ are similar to the previous case with exception that these 
curves show that the initial value of $\sigma$ is maintained for a short 
initial period of time. This period of time corresponds to
the stage of initial increase of the semi-major axis. 
\begin{figure}
\vspace{8cm}\includegraphics{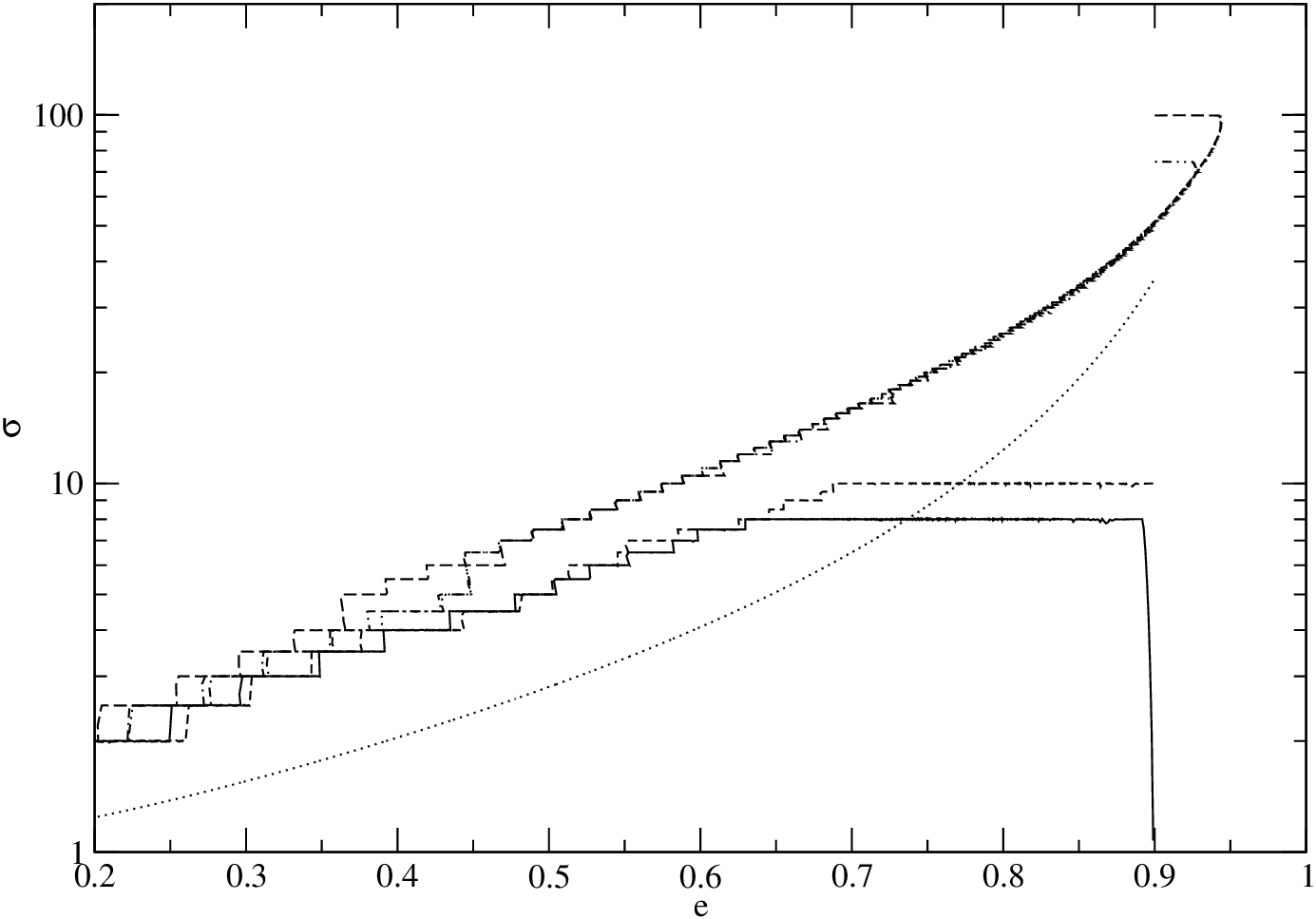}
\caption{The same as Fig 7. and Fig. 8, but the dependence of $\sigma (t)$ on eccentricity $e(t)$ is shown. Note that the
curves corresponding to $\sigma_{in}=50$, $75$ and $100$ (the long-dashed, dot-dashed and dot-dot-dashed curves, respectively)
are close to each other and practically coincide at large values of $e$.
The dotted curve represent the analytic solution $\sigma_{H}(e)$ corresponding to the case of the standard 
Navier Stokes viscosity (see equation \ref{eqno 61}) \label{fig10}}\end{figure}
In figure \ref{fig10} we show the evolutionary tracks of our dynamical system on the plane $(e,\sigma)$. As it is seen from this figure the
late time evolution proceeds through a sequence of the spin-orbit resonances with decreasing resonance order $n$. The dotted curve
shows the analytic dependence  $\sigma_{H}(e)$ corresponding to the case of the standard 
viscosity. It is important to point out that all curves corresponding to $p=2$ have $\sigma > \sigma_{H}$ for, 
practically, all values of eccentricity. Thus, a delayed response of convective motions to the tidal forcing can lead
to a high state of rotation of the primary star. It is also very interesting to note that the dynamical system can evolve
in a state of spin-orbit resonance even when the eccentricity is relatively small $\sim 0.2$.  
\begin{figure}
\vspace{8cm}\includegraphics{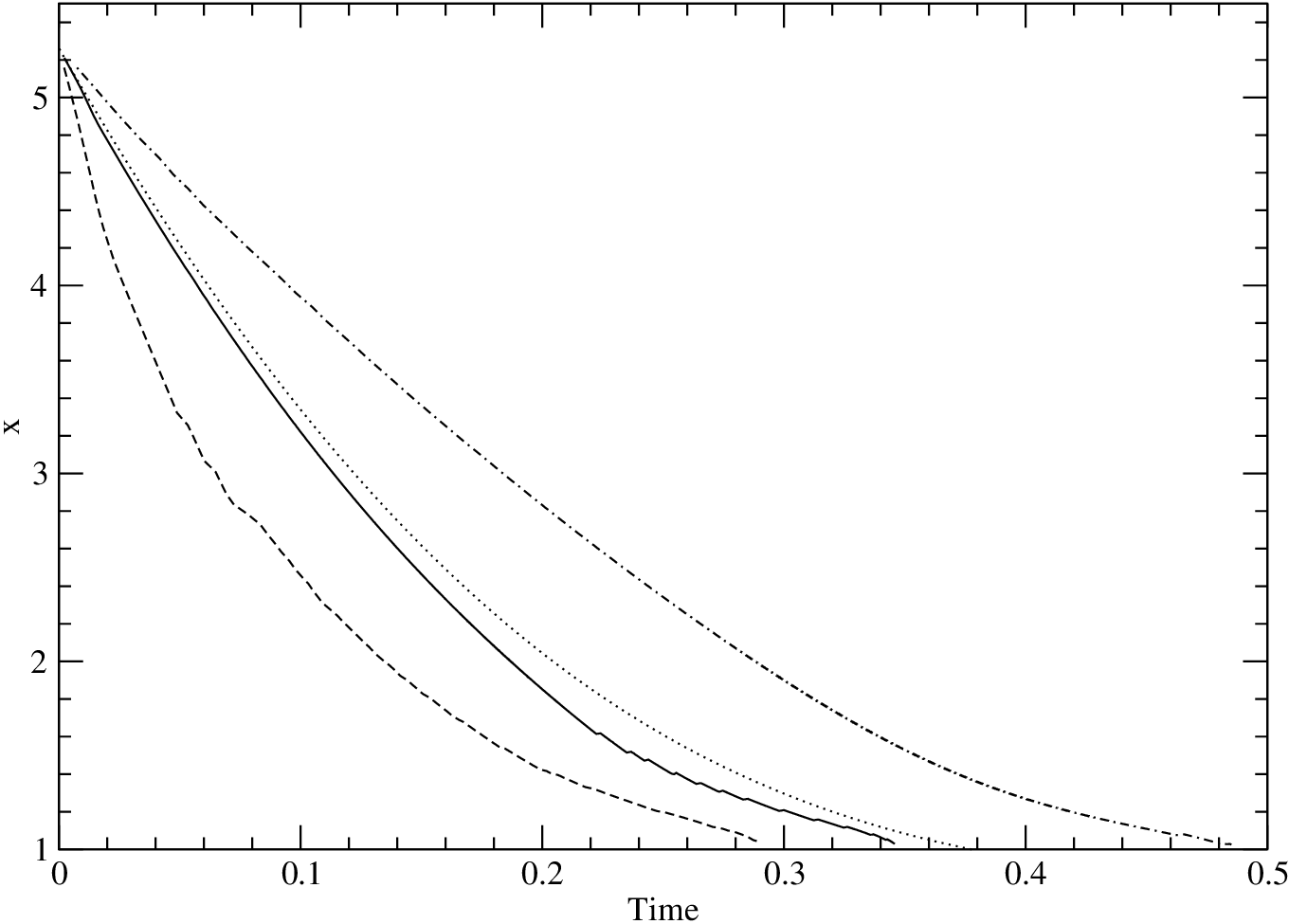}
\caption{The dependence of $x=a/a_{0}$ on time $\tilde t$ for different values of $p$. The solid, dashed and dot-dashed 
curves correspond to $p=2$, $1.5$, $2.5$, respectively.  $\sigma_{in}$ is set to zero. The dotted curve represents the 
analytic solution (\ref{eqno 73}).\label{fig11}}
\end{figure}
\begin{figure}
\vspace{8cm}\includegraphics{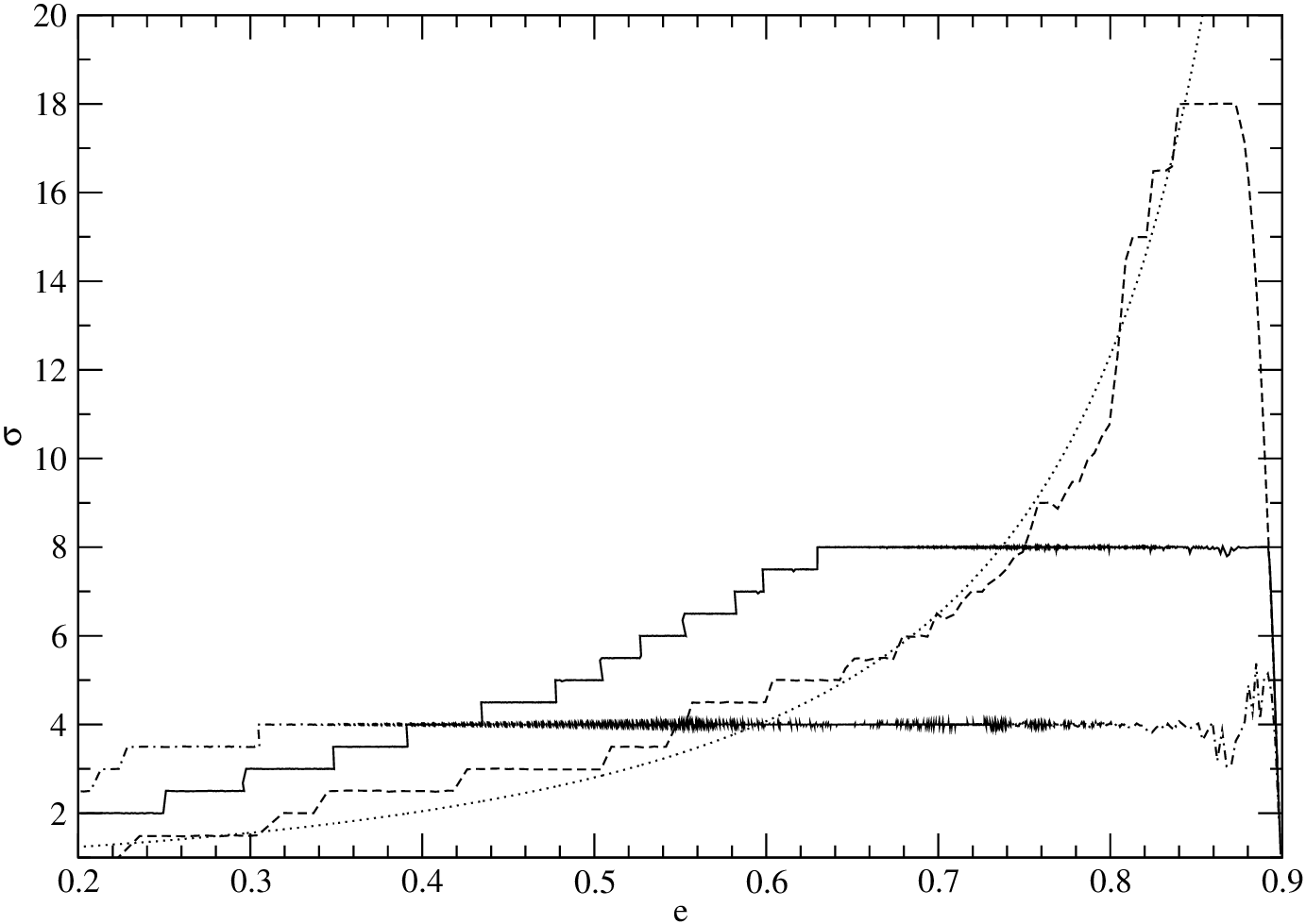}
\caption{The dependence of $\sigma (t)$ on eccentricity $e(t)$ 
is shown for $p=2$ (the solid curve), $p=1.5$ (the dashed curve) and
$p=2.5$ (the dot-dashed curve). The dotted curve represent the analytic solution $\sigma_{H}(e)$ corresponding to the case of 
the standard Navier Stokes viscosity (see equation \ref{eqno 61}).\label{fig12}}
\end{figure}

Finally, let us shortly discuss the case of $p\ne 2$. 
In figures \ref{fig11} and 
\ref{fig12}  we show the dependence of $x$ on time $\tilde t$ and
the evolutionary tracks on the plane $(e,\sigma)$ for $p=1.5$ and $p=2.5$ in comparison with the previous case of $p=2$.
The initial value of the rotational parameter $\sigma_{in}$ is set to zero. From figure 11 it is seen that the case $p=1.5$ 
($p=2.5$) evolves faster (slower) than the case $p=2$ with respect to the time $\tilde t$ toward the low eccentricity state.
This effect is similar to that was found in the case of small $p < 1$ considered above (see figure 5). Figure 12 shows that
the effect of evolution through a sequence of spin-orbit resonances remains valid for the different values of $p > 1$. The
resonances are less prominent for the $p=1.5$ case, and therefore the corresponding curve follows the curve representing the
standard Navier Stokes viscosity case on a large scale. On the other hand, the resonances corresponding to $p=2.5$ case 
are very strong, and the corresponding dynamical system evolves in $\sigma=4$ resonance until a small 
eccentricity $e\sim 0.3$ is reached.

\section{Discussion and conclusions} \label{CONC}
In this Paper we  have developed  a new self-consistent formalism for the calculation of
the rate of energy and angular momentum transfer  
from an eccentric  binary orbit to a fully convective
primary planet or star. This is based  
on  Fourier expansion in time  of the tidal forcing potential
and a  normal mode expansion of the tidal response.
To undertake calculations of orbital
 evolution, we have considered only the  $(l=2)$ fundamental mode which dominates
the response when the primary does not rotate, or when any potentially
resonant inertial modes present when it 
rotates, have poor overlap with the forcing potential.

We have assumed that the dissipation of the tidal response 
is due to an effective viscosity
induced by convective turbulence. 
This is presumed to have an associated relaxation time or eddy turn
over time which causes a delocalization of the dissipation process in time.
It results in a weakening of the dissipation for tidal forcing at relative  frequencies
exceeding $1/t_c.$
We considered the case  when the frequency dependence of the dissipative response is
$ \propto 1/(1+(\omega_{m,k}t_{c})^{p})$, where
$\omega_{m,k}$ is the apparent frequency associated
with the tidal forcing as viewed in the frame corotating with the primary in detail.

We use the fact that the orbital frequency is significantly
smaller than the eigen-frequency of the
$(l=2)$ $f$ mode to 	
introduce  the equilibrium tide approximation.
We note that, in general, this
approximation is not equivalent to the classical  assumption
of a constant time lag  between tidal forcing
and response (eg. Alexander 1973, Hut 1981). However, we do show that they are 
equivalent
only for the standard Navier Stokes viscosity law with instantaneous action 
of the viscosity $(p=0)$. 

We determine  
the orbital and rotational 
evolution of the primary 
numerically and analytically assuming that the angular momentum associated with the
primary is much less
than the orbital angular momentum
for a range of values of $p.$
We found that
the evolution of the primary does not depend significantly on $T_{*} = \Omega t_c,$
$\Omega$ being the orbital frequency,  provided that $T_{*}
\gg 1$ and
a scaled time  $\tilde t \propto  T_{*}^{-p}t$ is used.

However,
the dependence on the parameter $p$  was found to be significant. 
In the case of small $p < 1$ the evolution is similar to the case 
of the standard  Navier Stokes viscosity $(p=0).$  When $p > 1$ and  $T_{*}
\gg 1,$
the system evolves through a sequence
of specific spin-orbit corotation resonances with $\Omega_{r}/\Omega=n/2$, where
$\Omega_{r}$ 
is the rotation frequency and $n$ is an integer.  
For $p=2$ we found an analytical expression for the evolution of semi-major axis
given by equation (\ref{eqno 73}).  

We  stress that  for $p > 1$ the primary rotation frequency  at
'pseudo-synchronisation' is larger
than  obtained in the
standard constant time lag approximation $(p=0)$
and that the primary can evolve in a state of
spin-orbit resonance even when the
eccentricity is small ($e\sim 0.2$).
This may have observational
consequences. In principal,
one could use observations of the rotation of a sufficiently bright star in a close eccentric
binary to test different models for
response of convective turbulence to the tidal forcing. 

It should be possible to extend our
 formalism in several respects. For example, it can be
generalised to take into account
binaries with misaligned directions of  orbital and rotational angular momenta (for the
constant time lag case such
a generalisation has already  been made (eg. Alexander 1973, Hut 1981)).

One can also extend the  use of  equations (\ref{eqno 24} - \ref{eqno 25}) to a
situation where the contribution of
other terms in the denominator 
(\ref{eqno 25}) are non negligible compared with  $\omega_{0}^{2}$
(as in fact  occurs for sufficiently high harmonics).
 Taking into account a combination of  
several orthogonal modes responding to the tidal forcing is also straightforward.
This could
generalise our formalism to include the lower order
$g$ modes in convectively stable stars. 

Instead of using the model frequency dependence
of the viscous dissipation (equation (\ref{eqno 55}))
one could hope to determine more
realistic  specifications from numerical simulation of  convective media
under a periodic forcing. 
Finally, one  should use a realistic model of a
planet or a star to find the
response of the structure to the tidal heating 
and solve numerically the equations of evolution of the primary structure together with
the equations of orbital evolution.

Now let us discuss a possible restriction of our formalism  due to the fact that the
orbit  may not be  strictly periodic. Note that
the binary period is changing with time as a result of  energy transfer   between the
primary and the orbit.
Let the change of the orbital period during  one 
orbital period  be
$\Delta P_{orb}$ and the corresponding change in the orbital frequency be
$\Delta \Omega =
-{\Delta P_{orb}\over P_{orb}}\Omega$. 
We see from equation (\ref{eqno 16})  that the mode amplitude $b_{m}$
is expressed as a sum of harmonic
terms,
each of them being proportional to $e^{i\omega_{k}t}$, where $\omega_{k}=k\Omega$.
When
the number $k$ is sufficiently large
$k > k_{max}$, the change in the frequency $\Delta \omega_{k}=k|\delta \Omega|$ is larger
than the difference between two neigboring
frequencies equal to $\Omega$. The expansion in Fourier series is clearly invalid for $k >
k_{max}$, where
\be k_{max}= \left |{P_{orb}\over \Delta P_{orb}}\right |.  \label{eqno d1} \ee
Now let us consider the resonance $k=k_{r}\approx \omega_{0}/\Omega$ which is found from
the condition that 
the denominator $R$ (equation (\ref{eqno 25})) is close to zero,
and the corresponding $b_{m,k_{r}}$ could
be large.
When $k_{r} > k_{max}$, the
corresponding term in series (\ref{eqno 16}) is not periodic,
and the corresponding amplitude may be
amplified with time. From the condition
$k_{r} > k_{max}$ we find
\be \omega_{0}\Delta P_{orb} > 1, \label{eqno d2}\ee
which, to within  a numerical factor coincides with a condition for stochastic growth of the
mode amplitude due to dynamic tides
found by IP.
When this condition is fulfilled and the energy transfer from the orbit to
the mode is larger than the energy 
transfer associated with the equilibrium tide, our approach is not valid.

Finally we comment that our results for the time to circularize
an orbit starting from large eccentricity, $t_{circ},$ 
can be summarised for the various
values of $p$ in the compact form

\be t_{circ} \approx t_{\nu}\tilde t = {5M_p \omega_0^4 a_0^8 \tilde t \over
3\pi (GMQ)^2 \tilde \gamma (1+q)} \label{TCIRC}.\ee

\noindent Using equation (\ref{eqno 57}) for $\tilde \gamma,$ this becomes
\be t_{circ} \approx {5M_p \omega_0^4 a_0^8 \tilde t \over
3\pi (GMQ)^2 \gamma_0 (1+q)} \left(t_c\sqrt{GM/a_0^3}\right)^p \label{TCIRC1}.\ee

\noindent The dissipation rate $\gamma_0$ is given in IP as $\gamma_0 \sim 0.1/t_c,$
inserting this into (\ref{TCIRC1}) and making some other reductions
we obtain

\be t_{circ} \approx {5 \over 3\pi} \left[{\omega_0^2R_p^3\over GM}\right]^2
\left[{M_p R_p^2 \over Q^2}\right] {q^2 a_0^8\over  R_p^8}
\left( {10 t_c \tilde t \over  1+q } \right) 
\left(t_c\sqrt{GM/a_0^3}\right)^p \label{TCIRC2}.\ee

\noindent The product of the quantities in square brackets in (\ref{TCIRC2})
can be estimated for a Jupiter mass protoplanet in the late stages of its evolution
from data given in IP to be $  \sim 8.$ Similarly IP gives $t_c > 1yr$.
Using these values and the representative value $\tilde t =0.2,$ 
we find for $q= 10^{-3}$ that

\be t_{circ} > 10^{11} \left( {7\times 10^9 cm \over R_p}\right)^8
\left( {a_0\over 0.05AU}\right)^8\left(t_c\sqrt{GM/a_0^3}\right)^p yr  \ee

This rather long time scale  confirms the finding of IP
on the basis of an impulsive treatment of orbits with very high
eccentricity that equilibrium tides associated with 
the fundamental mode may not 
account for the circularization of the orbits of the  
recently discovered extra-solar planets with 
orbital periods of a few days $(a_0 = 0.05AU)$  even when $p=0.$

Use  of more plausible values, say, $p > 1$ would
give circularization times well beyond 
a life time of the planetary systems $\sim$ a few Gyrs.
This may indicate that dynamic tides in the  protoplanet
(and possibly also the star for higher mass protoplanets ) and tidal inflation 
arising from dissipation  in the
protoplanet is important
(see, eg. IP and references
therein). 
Note, however, that  the physics of 
convective turbulence is very poorly understood and  the 
validity of the mixing length theory used is an issue.
Also the contribution of
the spectrum of  inertial modes to the tide response
 may significantly shorten the circularization time scale
(see eg. Ogilvie $\&$ Lin 2003).

\section*{Acknowledgements}
The authors acknowledge support from PPARC through
research grant  PPA/G/02001/00486

\appendix
\section[]{Approximate analytic form   for the tidal Fourier coefficients }

As we discuss above the Fourier coefficients of 
the dimensionless tidal forcing amplitude $\phi_{m,k}$ can be
expressed in terms of two real quantities  $\phi_{m,k}={1\over 2}(\alpha_{m,k}-\beta_{m,k})$, where   
\be \alpha_{m,k}={1\over \pi}\int^{2\pi}_{0}d\tau {\cos(m\Phi(\tau))\cos(k\tau)\over \tilde D(\tau)^{3}}, \quad
\beta_{m,k}={1\over \pi}\int^{2\pi}_{0}d\tau {\sin(m\Phi(\tau))\sin(k\tau)\over \tilde D(\tau)^{3}}. \label{A3} \ee

Furthermore it is sufficient to consider only the case of $m=0,2$ and positive $k$, and for this case 
we approximately have 
$\alpha_{m,k}\approx \beta_{m,k}$.
Accordingly  we find approximate expressions only for for 
$\alpha_{2,k}, {\rm{and}}  \  \alpha_{0,k}$ hereafter. 

The integrals  given in (\ref{A3}) can be represented in a more useful form
with respect to the eccentric anomaly $\xi$ such that
\be \alpha_{2,k}={1\over \pi}\int^{2\pi}_{0} d\tau {R(e,\cos(\xi))\over {(1-e\cos(\xi))}^5}\cos(k\tau), \quad
\alpha_{0,k}={1\over \pi}\int^{2\pi}_{0} d\tau {1\over {(1-e\cos(\xi))}^3}\cos(k\tau),  \label{A4} \ee
where         
\be R(e,\cos(\xi))=2e^2-1-2e\cos(\xi)+(2-e^2)\cos^2(\xi). \label{A5} \ee

The integrals (\ref{A4}) cannot be evaluated analytically.  Hence, in order to obtain
an analytic representation, some reasonable 
approximations are needed. Often the integrands are
developed in powers of the eccentricity $e$ and integrated term by term.
However, the resulting series converge very slowly and  are not convenient for
analytic calculations involving 
orbits with  even a moderate eccentricity  $e\sim 0.5.$ 
Here we  use another approach based on the fact that the integrals are mainly determined by
the region of integration near  periastron $\xi \approx 0$. Accordingly, we expand the quantities 
$R(e,\cos(\xi)), \tilde D, (\xi - e\sin(\xi))$ in a  power series in $\xi.$ 
Substituting these series in  (\ref{A4}) and truncating them, we
get integrals which can be easily evaluated in principle.
This approach gives good results
in the limit of highly eccentric orbits $e \rightarrow 1$,  and also for 
eccentricities as small as $0.2$ and sufficiently large values of $k$.

At first, let us discuss the evaluation of $\alpha_{2,k}$.
In order to obtain an approximate value of $\alpha_{2,k}$
we expand $\tau(\xi)$  and $\tilde D(\xi)$ in powers of $\xi$ keeping terms up to 
third order. Then we may write
\be \tau\approx \sqrt{{2(1-e)^{3}\over e}}(x+{x^3\over 3}), \label{A6}\ee
\be \tilde D \approx (1-e)(1+x^2),\label{A7} \ee  
with  $x$  being a scaled form of $\xi$ defined such that 
\be x=\sqrt{{e\over 2(1-e)}}\xi. \label{A8} \ee
For an accurate approximation to $\alpha_{2,k}$,
the quantity $R(e,\cos(\xi))$ must be expanded  up to at least  
fourth   order so that
\be R(e,\cos(\xi))\approx (1-e)^{2}(1-{2(2+e)\over e}x^2+{8-e-4e^2\over 3e^3}x^4). \label{A9}\ee  
Substituting equations (\ref{A6} - \ref{A9})  into  equation (\ref{A4}) we get
\be \alpha_{2,k}\approx {1\over \pi(1-e)^3}\int^{2\pi}_{0}d\tau {\cos (k\tau)\over (1+x^2)^5}
(1-{2(2+e)\over e}x^2+{8-e-4e^2\over 3e^3}x^4). \label{A10} \ee
Finally, we need  to express $x$ in terms of $\tau$. For that, formally, one should solve the cubic equation 
(\ref{A6}), but this would lead to significant technical difficulties.
Instead we  again make the assumption  that the integral is
mainly determined by the values  of the integrand 
at  small values of $x$ and therefore  $\tau.$  
This enables us to make use of the following
approximate solution of equation (\ref{A6}) 
which is correct to fifth order in these quantities
and which also gives a linear relation between $x$ and $\tau$ for
large $x$ in the form   
\be x\approx \tilde \tau-{\tilde \tau^3\over 3(1+\tilde \tau^2)}, \label{A11}\ee
where we  adopt the re-scaled time 
\be \tilde \tau={\tau\over \lambda}, \quad \lambda=\sqrt{{2(1-e)^3\over e}},  \label{A12} \ee
Using equation (\ref{A11}) to substitute for $x$ in   the integral (\ref{A10}) we get
\be \alpha_{2,k} \approx {2 \over \pi \lambda e}\left (J_{0}
-{2(2+e)\over e}J_{1}+{8-e-4e^2\over 3e^3}J_{2}\right ), \label{A13}\ee
where
\be J_{i}=\int^{\infty}_{-\infty}d\tilde \tau 
{(1+\tilde \tau^{2})^{5-2i}\tilde \tau^{2i}(1+{2\over 3}\tilde \tau^2)^{2i}\over (1+2\tilde \tau^2+{\tilde \tau^4\over 3})^{5}}
\cos (k\lambda \tilde \tau). \label{A14} \ee
Here, again making use of  the assumption 
that the integral is mainly determined 
in a small interval near $\tilde \tau  = 0,$
and therefore to a good approximation independent of the interval
it is taken over, as long as that  is big enough,
we have changed the limits of integration to $\pm \infty.$

Note that we have also  neglected terms proportional to $\tau^{6}$ 
in the denumerator of the integrals (\ref{A14}). These terms
may easily be taken into account in principle but  do not
change significantly the results.
The integrals (\ref{A14})) can be easily evaluated  using 
the theory of functions of a complex variable.
They are determined by two poles of fifth order 
in the complex $\tilde \tau$  upper half-plane located at 
$\tilde \tau =i\sqrt {(3\mp \sqrt(6))}$.
The $J_{i}$ so found  can be represented in the form
\be J_{i}=e^{-y_{+}\lambda k}\sum^{n=4}_{n=0}a^{n}_{i}{(\lambda k)}^{n}
+e^{-y_{-}\lambda k}\sum^{n=4}_{n=0}b^{n}_{i}{(\lambda k)}^{n}, \label{A16}  \ee 
where $y_{\pm}=\sqrt {(3\mp \sqrt(6))}$,
and the coefficients $a^{n}_{i}$ and $b^{n}_{i}$
are given in Tables \ref{table1} and \ref{table2} respectively.

\begin{table*}
\begin{center}
\begin{tabular}{|l|l|l|l|l|l|l|l|l|l|l|l|l|l|}\hline\hline
       &     &     &     &     &        &        &           
& \\
$a^{n}_{i}$ &$ 0$& $1 $&$ 2 $&$ 3$& $4$ \\
       &     &     &     &     &        &        &             
&  \\
 \hline
 \hline
$0$&   $ 3.57\cdot 10^{-2}$& $2.05\cdot 10^{-2}$ & $ 2.91\cdot 10^{-2}$ & $ 1.34\cdot 10^{-3}$& $1.83\cdot 10^{-5}$ \\

$1$&   $ 6.76\cdot 10^{-2}$& $-1.68\cdot 10^{-2}   $&  $-1.13\cdot 10^{-2}   $&  $ -9.74\cdot 10^{-4}  $&
 $-2\cdot 10^{-5}$\\

$2$&   $ -9.97\cdot 10^{-3}$& $ -2.31\cdot 10^{-2} $&  $ 2.115 \cdot 10^{-5}$&  $ 5.25 \cdot 10^{-4} $& 

$ 2.18\cdot 10^{-5}$\\
\hline
\end{tabular}
\end{center} 
\caption{ \label{table1} The coefficients $a^{n}_{i}$ determining (together with $b^{n}_{i}$) the values of integrals
$J_{i}$ given by equation (\ref{A14}).  Each row is labelled by
 the value of $i$, and   each column  by the value of $n$
 (see equation (\ref{A16})).}
\end{table*}

\begin{table*}
\begin{center}
\begin{tabular}{|l|l|l|l|l|l|l|l|l|l|l|l|l|l|}\hline\hline
       &     &     &     &     &        &        &                
& \\
$b^{n}_{i}$ &$ 0$& $1 $&$ 2 $&$ 3$& $4$\\
       &     &     &     &     &        &        &               
&  \\
 \hline
 \hline
$0$ &  $-4.77\cdot 10^{-2}$& $-5.17\cdot 10^{-2}$& $-6.12\cdot 10^{-3}$&  $ 1.33 \cdot 10^{-2}$& $ 5.64\cdot 10^{-3} $\\

$1$&   $-9.25\cdot 10^{-3}   $& $ 4.54\cdot 10^{-2} $&  $ 7.04\cdot 10 ^{-2}   $&  $ 2.29\cdot 10^{-2}   $& 
$-1.08\cdot 10^{-2} $\\

$2$  & $ 5.84\cdot 10^{-2} $& $ 1.52\cdot 10^{-2} $ & $ 6.405\cdot 10{-2}   $ & $-1.36\cdot 10^{-1} $& 

$ 2.05\cdot 10^{-2} $\\
\hline
\end{tabular}
\end{center} 
\caption{ \label{table2} Same as table \ref{table1}  but for the coefficients $b^{n}_{i}$.}
\end{table*}

\begin{table*}
\begin{center}
\begin{tabular}{|l|l|l|l|l|l|l|l|l|l|l|l|l|l|}\hline\hline
 &     &     &     &     &
 & \\
 $n$& $ 0$& $1 $&$ 2 $ &$n$  &$ 0$& $1 $&$ 2 $\\
 &     &     &     &     &
 &  \\
 \hline
 \hline
 $c_{n}$&     $ 0.499      $& $ 0.141     $ & $ 6.38\cdot 10^{-3} $
 & $d_{n}$ &   $-3.81\cdot 10^{-2}   $& $ 0.141 $&  $ 0.199 $\\
 \hline
 \end{tabular}
 \end{center}
 \caption{ \label{table3} Coefficients $c_{n}$ and $d_{n}$ 
 determining the value of the integral (\ref{A18}). }
 \end{table*}

\newpage

We evaluate $\alpha_{0,k}$ by exactly  the same method. 
Doing this we get
\be \alpha_{0,k}\approx {2 \over \pi \lambda e} I, \label{A17}  \ee
where
\be I=\int^{\infty}_{-\infty}d\tilde \tau {{(1+{\tilde \tau}^2)}^{3}
\cos {(\lambda k \tilde \tau)}
\over (1+2\tilde \tau^2+{\tilde \tau^4\over 3})^{3}}. \label{A18} \ee
The integral (\ref{A18}) is determined by the behaviour at  two  poles of third order 
located at $\tilde \tau =iy_{\pm}$ in the upper complex half-plane and
can be expressed as
\be I=e^{-y_{+}\lambda k}\sum^{n=2}_{n=0}c_{n}{(\lambda k)}^{n} +
e^{-y_{-}\lambda k}\sum^{n=2}_{n=0}d_{n}{(\lambda k)}^{n}\ee
where the coefficients $c_n$ and $d_n$ are given in table \ref{table3} .
 
\bsp

\label{lastpage}

\end{document}